\documentclass[fleqn,usenatbib]{mnras}

\usepackage[T1]{fontenc}
\usepackage{ae,aecompl}

\usepackage{graphicx}	
\usepackage{amsmath}	
\usepackage{amssymb}	
\usepackage{courier}
\usepackage{xcolor}
\newcommand\fz{f_z}
\newcommand\error{\varepsilon}
\newcommand\be{\begin{equation}}
\newcommand\ee{\end{equation}}

\title[Stellar DF and the Local Vertical Potential]{The Stellar Distribution Function and Local Vertical Potential from Gaia DR2}

\author[H. Li et al.]{
Haochuan Li\thanks{E-mail: haochuan.li@queensu.ca}
and Lawrence M. Widrow\thanks{E-mail: widrow@queensu.ca}
\\
Department of Physics, Engineering Physics and Astronomy, Queen's University, Kingston, K7L 3X5, Canada
}

\date{Accepted 2021 February 22; Received 2021 February 20; In original form 2021 January 13}

\pubyear{2021}

\begin{document}
\label{firstpage}
\pagerange{\pageref{firstpage}--\pageref{lastpage}}
\maketitle

\begin{abstract}
We develop a novel method to simultaneously determine the vertical potential, force and stellar $z-v_z$ phase space distribution function (DF) in our local patch of the Galaxy. We assume that the Solar Neighborhood can be treated as a one-dimensional system in dynamical equilibrium and directly fit the number density in the $z-v_z$ plane to what we call the Rational Linear DF (RLDF) model. This model can be regarded as a continuous sum of isothermal DFs though it has only one more parameter than the isothermal model. We apply our method to a sample of giant stars from Gaia Data Release 2 and show that the RLDF provides an excellent fit to the data. The well-known phase space spiral emerges in the residual map of the $z-v_z$ plane. We use the best-fit potential to plot the residuals in terms of the frequency and angle of vertical oscillations and show that the spiral maps into a straight line.
From its slope, we estimate that the phase spirals were generated by a perturbation $\sim540$ Myr years ago. We also determine the differential surface density as a function of vertical velocity dispersion, a.k.a. the vertical temperature distribution. The result is qualitatively similar to what was previously found for SDSS/SEGUE G dwarfs. Finally, we address parameter degeneracies and the validity of the 1D approximation. Particularly, the mid-plane density derived from a cold sub-sample, where the 1D approximation is more secure, is closer to literature values than that derived from the sample as a whole.
\end{abstract}



\section{Introduction}\label{intro}

At the Sun's position in the Galaxy, the vertical force, that is, the gravitational force perpendicular to the Galactic plane, is dominated by baryons near the plane and dark matter above $1\,{\rm kpc}$. It therefore provides a powerful constraint on mass models of the Galaxy and helps break the disc-halo degeneracy. Moreover, when combined with measurements of the rotation curve near the Sun and the local baryon density, the vertical force yields an estimate for the local density of dark matter (see e.g. \citealt{read2014,deSalas2020}).

Stellar dynamics has been used to probe the vertical force since the pioneering work of 
\citet{Jeans1916a, Jeans1916b, Jeans1922, Kapteyn1922}, and \citet{Oort1932}. These studies lay the groundwork for early analyses by introducing three key assumptions for the vicinity of the Sun: (1) Stars behave like an incompressible fluid in six-dimensional phase space as described by the Collisionless Boltzmann Equation (CBE). (2) Near the mid-plane, the vertical force can be approximated by a function of $z$ while the stellar distribution function (DF) can approximated by a function of $z$ and $v_z$. Here, $z$ is the position relative to the mid-plane and $v_z$ is the velocity component perpendicular to the mid-plane. (3) Stars are in dynamical equilibrium with the gravitational potential. These assumptions together imply that the DF $f_z(z,v_z)$ depends on $z$ and $v_z$ through the vertical energy $E_z$, a result that follows from the Jeans Theorem. 
Alternatively, one can derive the vertical Jeans Equation, that is, the $v_z$ moment of the CBE, under the assumption of planar symmetry. The Jeans Theorem and the vertical Jeans Equation lead to various relations among the number density, velocity dispersion, and DF that can be used to estimate the vertical force and potential from kinematic data \citep{flynn1994, holmberg2000, holmberg2004, Zhang2013, Sivertsson2018, Buch2019, Guo2020}.

The mean field assumption is justified from the fact that the two-body relaxation time of stars in the disc is much greater than the age of the Galaxy \citep{GalacticDynamics}. As for the second assumption, which we refer to as the 1D approximation, radial variations become important as one moves further from the mid-plane with corrections reaching 10-20\% for $z>1\,{\rm kpc}$. They are also more important for stellar populations with higher radial velocity dispersion.
See \citealt{read2014} for a review. Note that a number of authors, including \citealt{bovy2013} and \citet{piffl2014} have devised methods to infer the vertical potential and force within the context of global models for the gravitational potential and three-integral models for the stellar DF. Finally, there is the assumption that the disc is in dynamical equilibrium. Oort looked for, but did not find, systematic motions in the direction perpendicular to the Galactic plane, and emphasized that this result lends ``support to the assumption that in the $z$-direction the stars are thoroughly mixed'' \citep{Oort1932}. However, we now know that the the disc has motions perpendicular to the plane, the most prominent being those associated with warping of the disc at large Galactocentric radii (see reviews by \citealt{Binney1992} and \citealt{kalberla2009}). Furthermore, there is evidence that the warp extends into the Solar Neighbourhood \citep{schonrich2018}. 
In addition, vertical bulk motions in disc stars and asymmetric variations in the number counts about the Galactic mid-plane have been identified by several groups using different surveys \citep{widrow2012, yanny2013, williams2013, carlin2013, bennett2019, salomon2020}. Perhaps the most striking indication of vertical disequilibrium comes from the phase spirals uncovered by \citet{antoja2018}. They can be seen in number counts as well as mean azimuthal and radial velocities across the $z-v_z$ plane and provide compelling evidence that the disc in the Solar Neighbourhood is not fully mixed \citep{antoja2018, binney2018, darling2019a, darling2019b, blandhawthorn2019, laporte2019, Li2020, Li2021}. Taken together, these observations bring into question the equilibrium assumption that is often made in attempts to measure the vertical force and local dark matter density \citep{banik2017,salomon2020} and call for new fitting methods that explicitly include disequilibrium features.

In this paper, we outline a method that represents the first step towards this goal and apply it to a sample of giants from \textit{Gaia}'s Second Data Release (GDR2). For this first step, we retain the three assumptions described above while simultaneously modelling the potential and DF directly in the full $z-v_z$ phase space. Specifically, we compare star counts in the $z-v_z$ plane, $N(z,\,v_z)$, with predictions from a model in which the DF and potential are described by simple parametric functions of the phase space variables. The best-fit potential is the one in which contours of constant $E_z$ come closest to contours of constant $N$. The method has the advantage of working directly with the DF and the potential. Furthermore, evidence of disequilibrium, such as the phase spirals, emerge in the residuals of the equilibrium model. By contrast, $n(z)$ (the $z$ distribution of stars) and $\fz(E_z)$ can hide manifestations of disequilibrium since they are constructed by integrating out one of the phase space coordinates ($v_z$ for $n(z)$ and an angle variable in the $z-v_z$ plane for $\fz$). As an extension of the model, we map the residuals into frequency-angle space via the best-fit potential and use it to infer the time when the disc was perturbed. In a subsequent paper, we will explore models where the phase spiral, underlying equilibrium DF, and potential are fit simultaneously.

The use of number count contours in the $z-v_z$ plane was considered by \citet{Kuijken1989a} who dismissed it for two reasons. First, the limited number of stars with full kinematic measurements available at that time would have necessitated fairly coarse bins in the $z-v_z$ plane. Second, measurement errors were difficult to handle and model uncertainties were difficult to estimate. In this paper, we analyse data from GDR2, which contains over 6 million stars with complete position and velocity measurements \citep{gaiadr2_summary}. By \textit{Gaia}'s Third and Fourth Data Releases, the number of stars with 6D phase space measurements will swell by over two orders of magnitude. Thus, we have the opportunity to fit $z-v_z$ contours with a sufficiently fine grid. Moreover, Markov Chain Monte Carlo methods allow us to efficiently estimate uncertainties via Bayesian statistics. We are therefore well-positioned to address the obstacles described in \citet{Kuijken1989a}.

The outline of the paper is as follows: We present our fitting algorithm in Section \ref{AlgoIntro} and test it on mock data in Section \ref{MockDataTest}. In Section \ref{GDR2Reduction}, we describe the selection criteria for our sample of GDR2 giants. We then present the results derived from this sample in Section \ref{GDR2Results} and discuss them in Section \ref{Discussion}. We conclude in Section \ref{Conclusion} with a summary and thoughts on future directions for this line of research.

\section{The fitting algorithm}\label{AlgoIntro}

\subsection{Likelihood function}

Consider a sample of stars with fully determined positions and velocities that are selected according to their intrinsic properties and locations within the Galaxy. In what follows, we assume that the sample is complete. That is, all stars inside the region of the sample and with the chosen set of stellar properties are assumed to be included in the sample. Generally, the number density of stars in the $z-v_z$ plane predicted by a model DF $f({\bf x},\,{\bf v})$ is given by:
\begin{equation}\label{eq:nzv_general}
n(z,\,v_z)  = \int f({\bf x},\,{\bf v})\, S({\bf x}) \,dx\,dy\,dv_R\,dv_\phi
\end{equation}
where the geometric selection function $S({\bf x})$ is unity inside the sample volume ${\cal V}$ and zero outside. 
In this paper, we further assume that the system is in dynamical equilibrium, axisymmetric, and symmetric about the mid-plane of the Galaxy. By Jeans theorem, the DF depends on the positions and velocities through the integrals of motion. For an axisymmetric system, the component of the angular momentum along the symmetry axis $L_z$ and the total energy $E$ are both exact integrals, while the vertical structure of the disc is determined by the dependence of the DF on the vertical energy
\begin{equation}\label{eq:Ez_ori}
    E_z = \frac{1}{2}v_z^2 + \Psi(R,\,z) - \Psi(R,\,0)~,
\end{equation}
which is only approximately conserved.

As mentioned in Section \ref{intro}, we make two additional assumptions for the Solar Neighborhood: (1) the stellar distribution function (DF) can be approximated by some $f_z(z,v_z)$, and (2) the vertical force is only a function of $z$. In this way, we can define the vertical potential as
\begin{equation}\label{eq:VerticalPot}
    \psi(z) \equiv \Psi(R,\,z) - \Psi(R,\,0).
\end{equation}
The vertical energy is then
\begin{equation}\label{eq:Ez}
    E_z = \frac{1}{2}v_z^2 + \psi(z),
\end{equation}
which is an integral of motion. Therefore, Equation \ref{eq:nzv_general} implies that:
\begin{equation}\label{eq:nzv_simplify}
n(z,\,v_z)\propto {\cal G}(z)\,\fz(z,\,v_{z})~=~{\cal G}(z)\,\fz(E_z)
\end{equation}
where
\begin{equation}\label{eq:Gz}
{\cal G}(z)=\int S({\bf x})dxdy
\end{equation}
is essentially the area of the intersection of the sample volume and a horizontal plane at height $z$.

We sort stars into $z-v_z$ bins and define $O_i$ to be the observed star count in the $i$'th bin, centred on the point $(z_i,\,v_{z,i})$. Let $A_i$ be the area of the $i$-th bin in $z-v_z$ phase space so that it has dimensions of distance $\times$ velocity. The prediction for the star counts in the $i$-th bin
is found by multiplying Equation \ref{eq:nzv_simplify} by
$A_i$:
\begin{equation}\label{eq:ModelCounts}
M_i\propto A_i{\cal G}(z_i)\fz(E_{z,i})
\end{equation}
with the normalization condition
\begin{equation}\label{eq:Norm}
\sum_i M_i=N
\end{equation}
where $N$ is the total number of stars used in the analysis. Since there are always a few stars falling outside our $z-v_z$ grid, $N$ is generally a bit less than the total number of stars in the sample. The probability of a star falling into the $i$-th bin is given by $p_i=M_i/N$. Our likelihood function is then the product of the binomial distribution over all bins:
\begin{equation}\label{eq:likelihood}
    {\cal L} = \prod_{i} \frac{N!}{O_i!\,(N-O_i)!}p_i^{O_i}\left (1-p_i\right )^{N-O_i}
\end{equation}

\subsection{Gravitational potential}

In this work, we adopt a reparameterization of the potential introduced by \citet{Kuijken1989a, Kuijken1989b}:
\begin{equation}\label{eq:KGPotential}
    \psi(z) = {\omega_1}^2 D\left (\sqrt{z^2 + D^2} - D\right ) + \frac{1}{2}{\omega_2}^2z^2~.
\end{equation}
They identify the first term, which is quadratic near the mid-plane and linear for $z\gg D$, with the disc. In this interpretation, the disc has thickness $D$ and surface density $\Sigma_d = \omega_1^2 D/(2\pi G)$. Likewise, they identify the second term with an effective halo having a constant density $\rho_{\rm eff} = \omega_2^2/(4\pi G)$. The term ``effective halo" is used since the bulge and disc also contribute quadratic components to the potential. However for our purposes Equation~\ref{eq:KGPotential} is simply a convenient fitting formula for the potential. 

For an axisymmetric system, Poisson's equation is given by
\begin{equation}\label{eq:poisson}
\frac{\partial^2\Psi}{\partial z^2} + \frac{1}{R}\frac{\partial}{\partial R}
\left (R\frac{\partial\Psi}{\partial R}\right ) = 4\pi G \rho(R,\,z)  
\end{equation}
where $\rho$ is the total mass density. If we assume that the radial contribution
is independent of $z$, then the left-hand side of Equation \ref{eq:poisson} becomes
\begin{equation}
4\pi G\rho = 
\frac{\omega_1^2}{(1 + z^2/D^2)^{3/2}} + \omega_2^2 + 2\left (B^2 - A^2\right )
\end{equation}
where $A$ and $B$ are the Oort constants. The integrated surface density $\Sigma$ within a distance $z$ ($z>0$) from the mid-plane is given by 
\begin{equation}\label{eq:Sigmaz}
2\pi G\Sigma(z) = F_z(z) +  2\left(B^2 - A^2\right)z
\end{equation}
where 
\begin{equation}\label{eq:Forcez}
F_z(z) = \left(\frac{\omega_1^2}{\sqrt{1 + z^2/D^2}} + \omega_2^2\right)z 
\end{equation} 
is the magnitude of the vertical force at a distance $z$ from the mid-plane. In the limit $z\to 0$ we have
\begin{equation}\label{eq:rho0}
4\pi G\rho_0 = \omega_1^2 + \omega_2^2 + 2\left(B^2 - A^2\right)
\end{equation}
where $\rho_0$ is the total density at the mid-plane.

\subsection{Distribution function}

In the 1D approximation, Jeans Theorem implies that the DF is a function of $E_z$. A particularly simple choice is the well-known isothermal DF \citep{spitzer1942, camm1950},
\begin{equation}\label{eq:isothermal}
  \fz(E_z)\propto e^{-E_z/{\sigma_z}^2}~.
\end{equation}
which yields a vertical velocity dispersion $\sigma_z$ that is constant in $z$. However, it is well-known that $\sigma_z$ increases with $z$. This observation lends evidence to the idea of a thin disc-thick disc dichotomy \citep{Gilmore1983} (see \citealt{DiskReview} for a review). 
Numerous authors have modelled tracers of the vertical force by combining isothermal components with different $\sigma_z$. \citep{Bahcall1984,holmberg2000,holmberg2004,Flynn2006}.
This idea is at the heart of the \citet{bovy2013} analysis
of the vertical force as a function of $R$, which builds on earlier work by \citet{bovy2012a} and \citet{bovy2012b} where stars are divided into mono-abundance sub-populations. These sub-populations are approximately isothermal with a vertical velocity dispersion that varies smoothly with elemental abundance. 
In addition, one can consider full three-dimensional models for the DF and potential. For example, the DF can be factored into quasi-isothermal terms for in-plane and vertical motions as in \citep{kuijken1995, binney2010, binney2011, piffl2014}.

In this work, we introduce the Rational Linear Distribution Function (RLDF)
\begin{equation}\label{eq:RLDF}
\fz(E_z) = f_0 \left(1+\frac{E_z}{\alpha{\sigma_z}^2}\right)^{-\alpha}
\end{equation}
where $f_z d^3 {\bf x} dv_z$ is the number of stars in a volume element $d^3{\bf x}$ and velocity element $dv_z$. Such a DF neatly produces an increasing $\sigma_z$ as a function of $z$:
\begin{equation}\label{eq:RLDFdispersion}
\langle v_z^2\rangle^{1/2} 
=\sigma_z\sqrt{\frac{\alpha}{\alpha-3/2}
\left(1+\frac{\psi(z)}{\alpha\sigma_z^2}\right)}
\end{equation}
while the number density of stars as a function of $z$ is then given by
\begin{equation}\label{eq:NumDens}
\nu(z) = \int_{-\infty}^\infty f(E_z)dv_z = \nu_0
\left (1 + \frac{\psi(z)}{\alpha {\sigma_z}^2}\right )^{1/2-\alpha}
\end{equation}
where
\begin{equation}
\nu_0 = \sqrt{2\pi\alpha}\frac{\Gamma(\alpha-1/2)}
{\Gamma(\alpha)}\sigma_zf_0.
\end{equation}
Note that we require $\alpha>3/2$ for the system to have a finite velocity dispersion.

Although the RLDF is manifestly non-isothermal, it is closely related to the isothermal profile. 
First, in the limit $\alpha\to\infty$, Equations \ref{eq:RLDF}-\ref{eq:NumDens} reduce to those for the isothermal DF, namely Equation \,\ref{eq:isothermal} for the DF and
\begin{equation}
  \nu(z) \to \nu_0 e^{-\psi(z)/{\sigma_z}^2}~\mbox{and}~
  \langle v^2\rangle^{1/2} \to \sigma_z
\end{equation}
for the stellar density and velocity dispersion. 
Moreover, the RLDF can be written as a continuous superposition of isothermal DFs through the integral
\begin{equation}\label{eq:RLDF_decompose}
    f_z(E_z) = f_0 \int_0^\infty
    \frac{d\mu_z}{\mu_z}
    g\left (\sigma_z/\mu_z;\alpha\right)
      e^{-E_z/\mu_z^2}
\end{equation}
where
\begin{equation}
  g(x;\,\alpha) = \frac{2\alpha^\alpha}{\Gamma(\alpha)} x^{2\alpha} e^{-\alpha x^2}
\end{equation}
The differential DF is then
\begin{equation}\label{eq:frac_dens}
  \frac{df}{d\mu_z} = f_0 \frac{g\left(\sigma_z/\mu_z;\alpha\right)}{\mu_z} e^{-E_z/{\mu_z}^2}~.
\end{equation}
We integrate this expression over $z$ and $v_z$ to obtain
\begin{equation}\label{eq:frac_Sigma}
    \frac{d\Sigma}{d\mu_z} \propto g\left(\sigma_z/\mu_z;\alpha\right)
    \int_0^\infty dz e^{-\psi(z)/{\mu_z}^2}
\end{equation}
where $\frac{d\Sigma}{d\mu_z}\Delta\mu_z$ is the contribution to the surface density for stars from populations with velocity dispersion between $\mu_z$ and $\mu_z + \Delta\mu_z$ assuming that stellar populations are well mixed.

\subsection{Fitting procedure}

To summarize, the potential is defined by three parameters $\omega_1$, $\omega_2$, and $D$ through Equation \ref{eq:KGPotential}, while the DF is defined by $\sigma_z$ and $\alpha$ through Equation \ref{eq:RLDF}. The normalization factor $f_0$ is calculated from other parameters via Equation \ref{eq:Norm}. We fix the bin size to be $(\Delta z,\,\Delta v_z)=(40\,{\rm pc},2\,{\rm km/s}$). Equation \ref{eq:likelihood} gives the probability of the data given the model. We calculate the probability distribution function (PDF) of the model parameters given the data via Bayes theorem assuming
linear priors for $\omega_1$, $D$, $\omega_2$, $\sigma_z$ and $\ln\,\alpha$ (instead of $\alpha$) as listed in Table~\ref{tab:prior}. To do so, we using the Markov chain Monte Carlo (MCMC) sampler \textsc{emcee} \citep{emcee}. 

\begin{table}
    \centering
    \renewcommand{\arraystretch}{1.4}
    \begin{tabular}{cl}
    Parameter & Prior Range  \\
    \hline
    $\omega_1$ & $[10,80]\,{\rm km/s/kpc}$  \\
    $D$ & $[0.1,1.0]\,{\rm kpc}$  \\
    $\omega_2$ & $[0,50]\,{\rm km/s/kpc}$ \\
    $\sigma_z$ & $[0,40]\,{\rm km/s}$\\
    $\ln\,\alpha$ & $[\ln\,(3/2),5]$
    \end{tabular}  
    \caption{Prior ranges for all parameters. Note that we need $\alpha>3/2$ for a finite vertical velocity dispersion.}
    \label{tab:prior}
\end{table}

\section{Mock data tests}\label{MockDataTest}

\subsection{Data generation}

In this section, we test our method on two mock datasets. The first is drawn from an equilibrium model, which is constructed using the \textsc{GalactICS} code \citep{kuijken1995, widrow2005,deg2019}.
It comprises 200k stars in an annulus centered at the Solar radius $R_0=8.3\,{\rm kpc}$ with a half-width of $500\,{\rm pc}$. We assume a sample volume that does not depend on $z$ so that ${\cal G}(z)=1$. The particular \textsc{GalactICS} model is described in \citet{darling2019a} and yields a Milky Way-like system that comprises an exponential disc, a S\'{e}rsic bulge and an NFW halo \citep{nfw1996} with the following properties:
\begin{itemize}
    \item The disc has a radial scale length $R_d=2.8\,{\rm kpc}$ and a mass of $M_d \simeq 3.8\times 10^{10}\,M_\odot$;
    \item The bulge has a S\'{e}rsic index of $2$, a scale length of $700\,{\rm pc}$ and a mass of $M_b \simeq 1.3\times 10^{10}\,M_\odot$;
    \item The circular speed at the solar circle is $\simeq 255\,{\rm km/s}$;
    \item At the Solar Circle, the thickness of the disc is $\langle z^2\rangle^{1/2} = 320\,{\rm pc}$, and the vertical velocity dispersion is $\sigma_z = 17\,{\rm km/s}$;
    \item The disc has a Toomre Q-parameter $Q=1.5$ at $R=2.2R_d$.
\end{itemize}
Our second partially-perturbed dataset includes a phase spiral, which is qualitatively similar to the one discovered by \citet{antoja2018}. To generate the phase spiral, we
give $\simeq 10\%$ of the particles from the equilibrium dataset an impulsive "kick" of $20\,{\rm km/s}$ at $t=0$ in the $z$-direction. The velocities of the remaining 90\% of the particles are unchanged. We then evolve the system
for $500\,{\rm Myr}$ in the fixed potential of the equilibrium galaxy.
Thus, the phase spiral that emerges in the $z-v_z$ DF of the perturbed particles is axisymmetric and
purely kinematic. 
(For a discussion of the importance of self-gravity in the development of phase spirals, see \citealt{darling2019a}.)
Though highly idealized, the set-up serves as a useful illustration of our method.
For each of the two mock datasets, we only keep stars within $|z|<1.48\,{\rm kpc}$ and $|v_z|<80\,{\rm km/s}$ for fitting, as few stars reach beyond this region of $z-v_z$ phase space.

\subsection{Results}

\begin{figure}
	\includegraphics[width=\columnwidth]{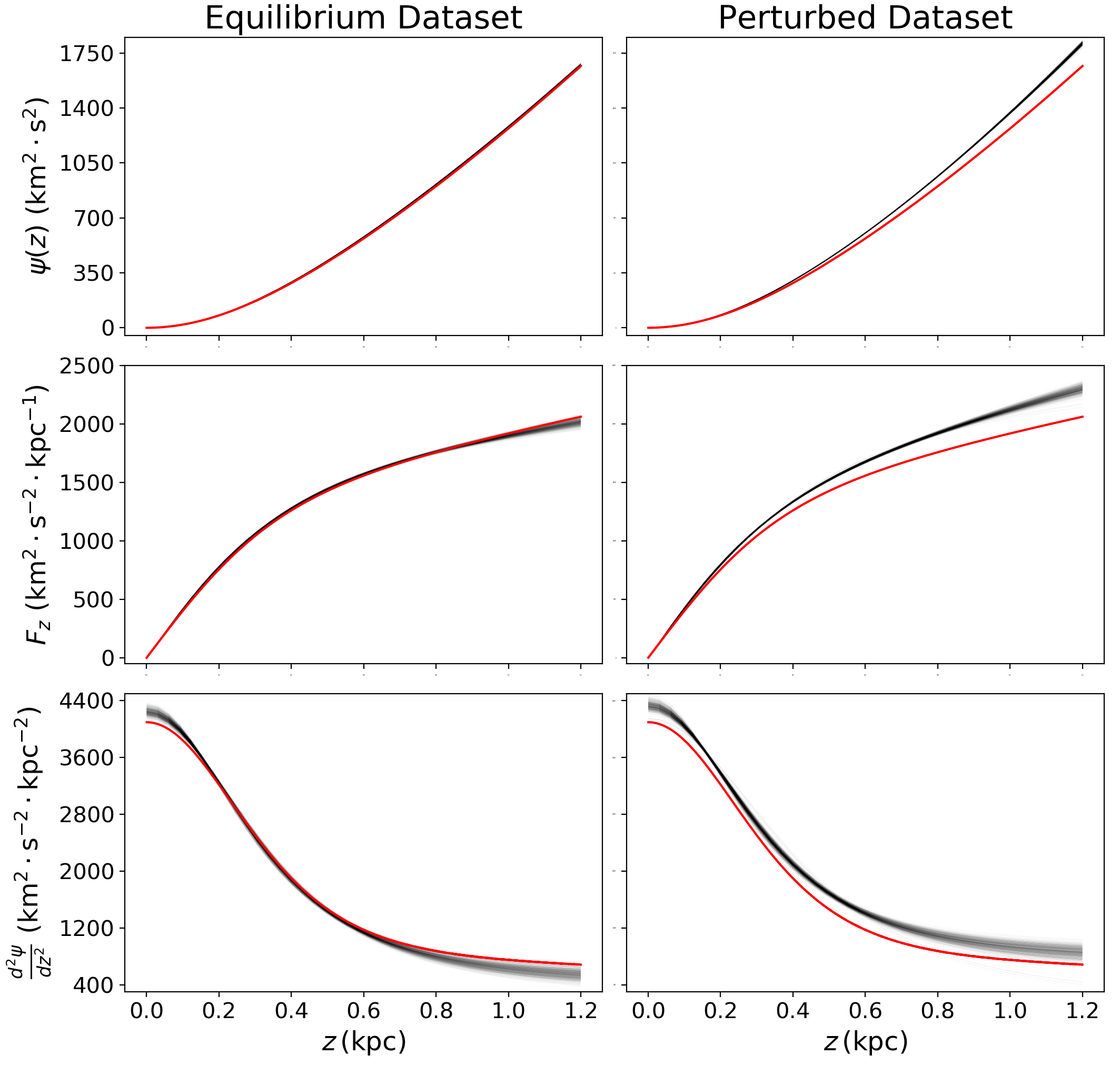}
        \caption{Profiles of $\psi(z)$ (top row), $F_z$ (middle row) and $d^2\psi/dz^2$ (bottom row) as derived from our fitting for the equilibrium mock dataset (left column) and the perturbed mock dataset (right column). We show 1000 random fitting realizations (grey lines) for each panel. The red lines show true profiles of these quantities as calculated at $R=R_0=8.3\,{\rm kpc}$.}
    \label{fig:profiles}
\end{figure}

In Figure \ref{fig:profiles}, we show the results for $\psi(z)$, $F_z = d\psi/dz$ and $d^2\psi/dz^2$ along with the true profiles as determined by the \textsc{GalactICS} code. In the case of $d^2\psi/dz^2$, the true profile is assumed to be given by $4\pi G\rho - 2\left(B^2 - A^2\right)$.
With the equilibrium dataset, the model does an excellent job of recovering the potential and force as can be seen in the left-hand panels of Figure \ref{fig:profiles}.
The model overestimates $d^2\psi/dz^2$ near the mid-plane by $\sim5\%$ and underestimates for $z\gtrsim 600\,{\rm pc}$ by $\simeq20\%$. The relatively poor fit at large $z$ is not surprising given the small number of tracers at these large distances above the mid-plane and the fact that one is attempting to extract a second derivative of the potential.

As one might expect, the potential and force are not recovered as well when the mock data includes a phase spiral. The model underestimates the potential and force while overestimating $d^2\psi/dz^2$. 

\begin{figure}
	\includegraphics[width=\columnwidth]{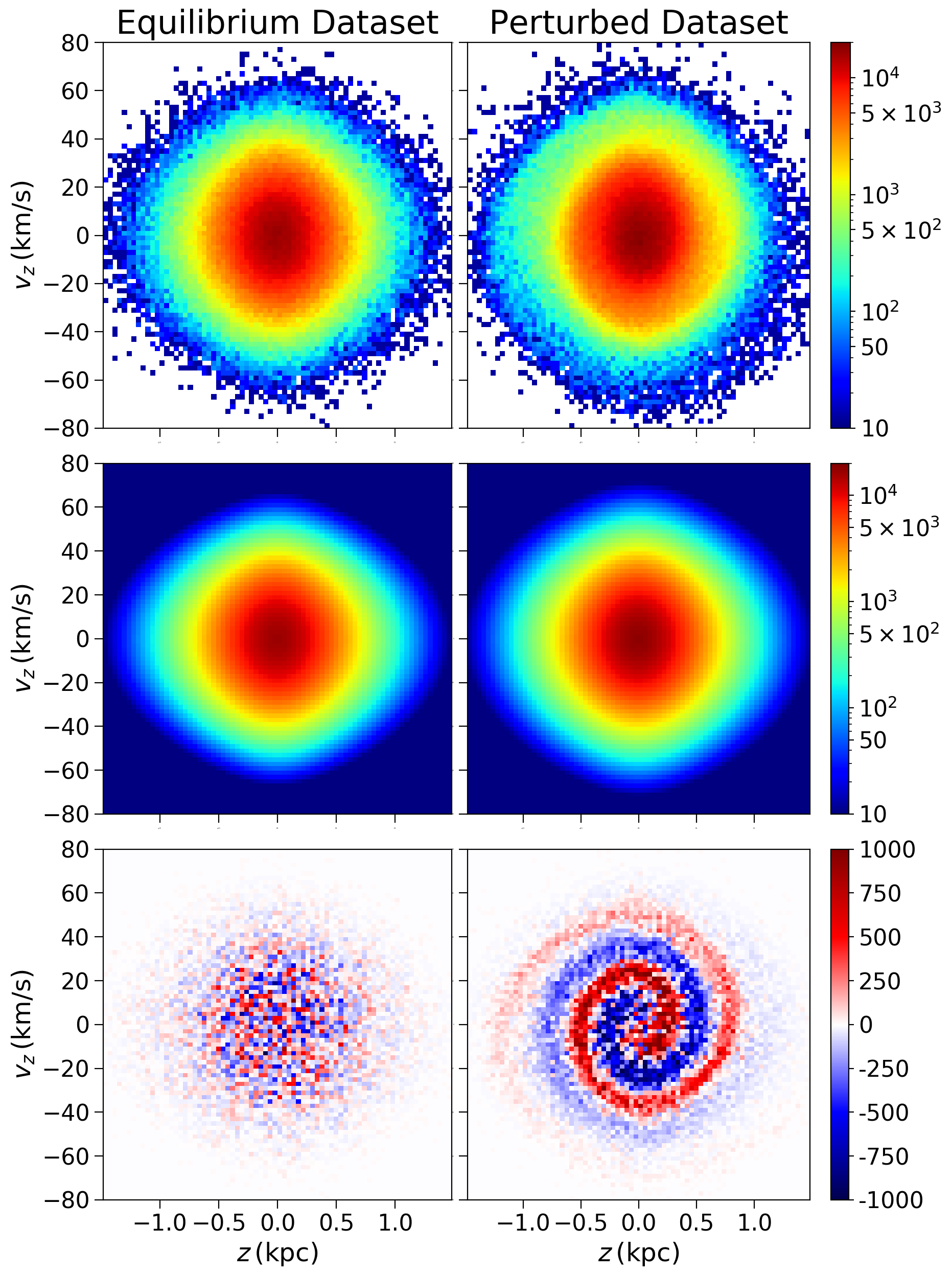}
         \caption{Number density in the $z-v_z$ plane for the equilibrium (left column) and the perturbed (right column) mock dataset derived from the data (upper row), the best-fit model (middle row) and the residual (lower row). The unit of the densities is ${\rm kpc^{-1}\cdot (km/s)^{-1}}$. Each row shares the same color bar.}
    \label{fig:mockresiduals}
\end{figure}

In Figure~\ref{fig:mockresiduals}, we show the actual and model-predicted number densities in the $z-v_z$ plane for both mock datasets as well as the residuals. For the equilibrium dataset, the model captures the structure of the number density map extremely well. In particular, the residuals are dominated by shot noise. On the other hand, the spirals dominate the residuals in the perturbed mock dataset.
 
\subsection{Residuals in frequency-angle space}\label{mock_AAV_fitting}

The phase spirals seen in Figure \ref{fig:mockresiduals} for the perturbed mock dataset are the result of phase mixing after the initial perturbation in an anharmonic vertical potential. A star with vertical energy $E_z$ orbits the $z-v_z$ plane with an angular frequency $\Omega(E_z) = 2\pi/T(E_z)$ where 
\begin{equation}\label{Period}
 T(E_z)=2\int_{-z_m}^{z_m}\frac{dz}{|v|}
       =2\sqrt{2}\int_{0}^{z_m}\frac{dz'}{\sqrt{E_z-\psi(z')}}
\end{equation}
is the orbital period and $z_m=z_m(E_z)$ is the maximum vertical excursion of a particle with energy $E_z$. Note that for $E_z=0$, $\Omega(0)\equiv\Omega_0=({\omega_1}^2 + {\omega_2}^2)^{1/2}$. The usual angle variable of angle-action coordinates $\theta$ \footnote{We define $\theta=0$ at $v_z=0,\,z>0$.} is then given by
\begin{equation}
    \theta(z,v_z)=-sgn(v_z)\cdot2\pi\cdot \tau/T(E_z)
\end{equation}
where $sgn$ is the sign function, $E_z$ follows Equation \ref{eq:Ez} and $\tau$ is the time it takes for a star at $z$ with energy $E_z$ to travel upwards to its maximum excursion.

The initial perturbation in our second mock dataset amounts to a displacement of the peak in the $z-v_z$ DF along the $v_z$ direction. Over time $t$, this peak is sheared due to variations in $\Omega(E_z)$ with $E_z$. So long as the particles evolve kinetically, the peak becomes a straight ridge 
in the $\Omega(E_z)-\theta$ plane as defined by the linear equation
\begin{equation}\label{eq:ridge}
  \theta = t\cdot\Omega(E_z) + \theta_0.
\end{equation}
plane \citep{antoja2018, binney2018, darling2019a}. In this equation, $\theta_0$ is the angle of the initial displacement. For our perturbed dataset, the true values are $t=500$ Myr and $\theta_0=-\pi/2$. 

\begin{figure}
	\includegraphics[width=\columnwidth]{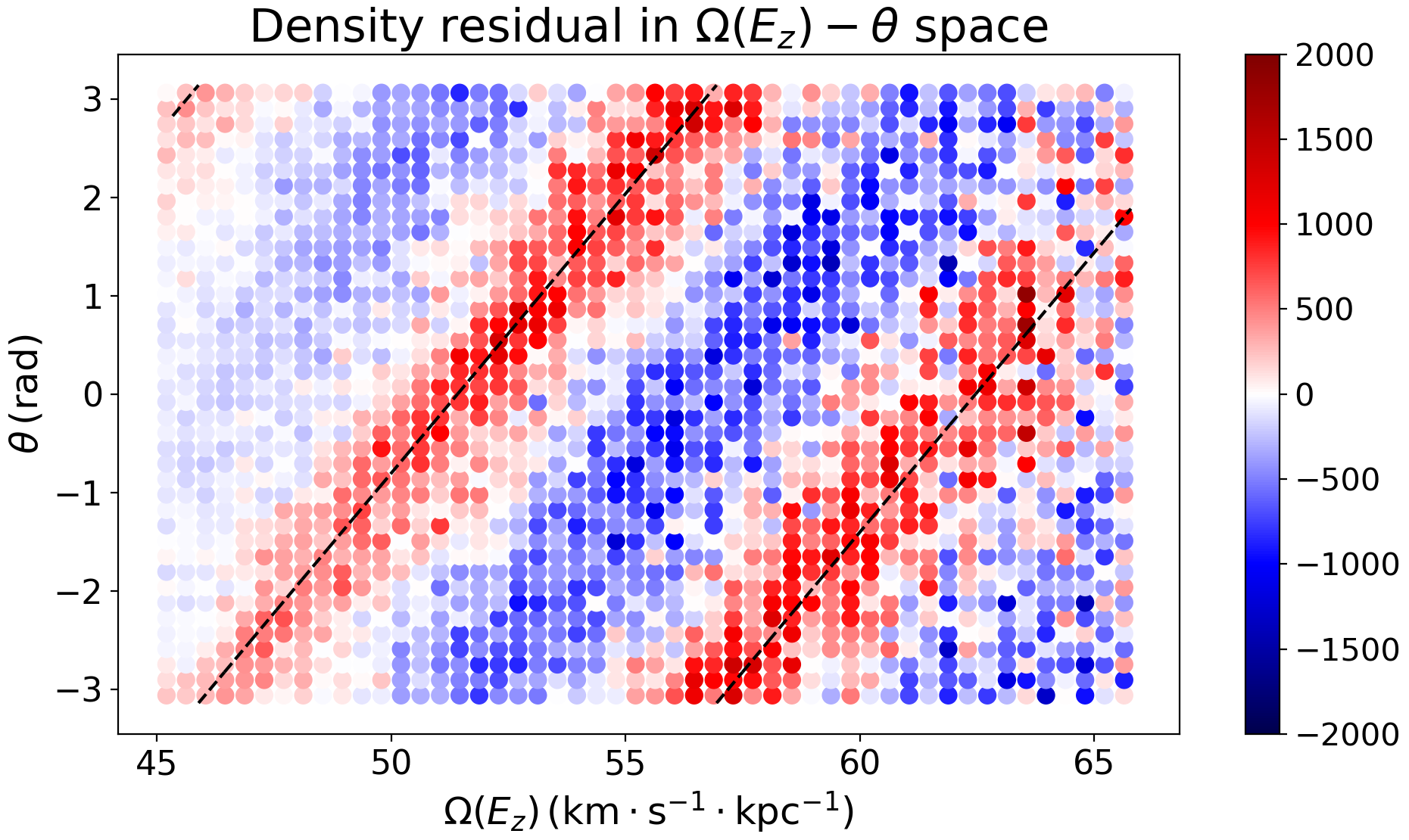}
    \caption{Residual map for the perturbed mock dataset in $\Omega(E_z)-\theta$ space. The solid black lines correspond to a perturbation age of $t=555.9\,{\rm Myr}$ and a phase shift of $\theta_0 = 2.2\,{\rm rad}$. The true values are $t=500\,{\rm Myr}$ and $\theta_0=-\pi/2\,{\rm rad}$. The unit of the residual is ${\rm kpc\cdot (km/s)^{-1}\cdot rad^{-1}}$.}
    \label{fig:mockAF}
\end{figure}

\begin{figure}
	\includegraphics[width=\columnwidth]{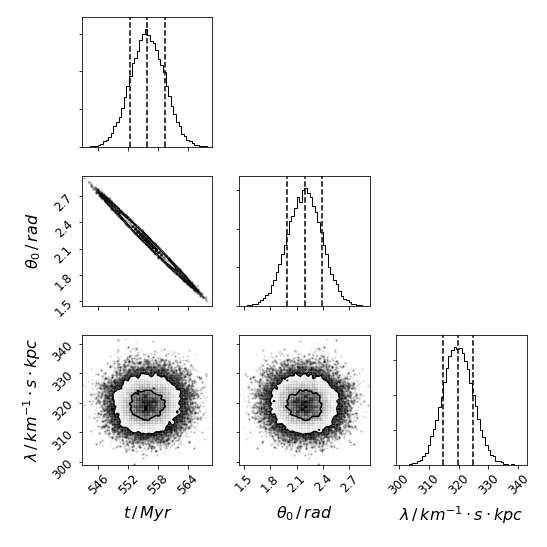}
    \caption{Parameter contour plot of the residual fitting in $\Omega(E_z)-\theta$ space.}
    \label{fig:mock_AAV_contour}
\end{figure}

In Figure \ref{fig:mockAF}, we map the residuals shown in Figure \ref{fig:mockresiduals} to the $\Omega(E_z)-\theta$ plane using the best-fit potential from our maximum likelihood analysis. The spirals do indeed become straight lines with the obvious wrap-around effect due to periodicity in $\theta$. To infer $t$ and $\theta_0$, we model the washboard in Figure \ref{fig:mockresiduals} with a Fourier series:
\begin{equation}
  R_F\left (\theta_i,\,\Omega_i;\,t,\theta_0\right ) =
  \sum_{n=1}^{n_{\rm max}}
  A_n\cos{\left[n\left (\theta_i-\Omega_i t-\theta_0\right )\right]}  
\end{equation}
where we expect the argument of the cosine to be constant along ridges and furrows. The subscript $i$ refers to the $i$'th bin of our $\Omega-\theta$ grid. We then take the log-likelihood function to be
\begin{equation}
\begin{split}
  \ln{\cal \tilde{L}}\left (t,\,\theta_0,\,\lambda\right )
 &  = -N_b\ln{\left (2\pi \lambda\right )} \\ -
  & \frac{1}{2\lambda^2} 
\sum_{i=1}^{N_b} 
\left[R_F(\theta_i,\,\Omega_i;\,t,\theta_0)-R(\theta_i,\,\Omega_i)\right]^2
\end{split}
\end{equation}
where $N_b$ is the number of $\Omega-\theta$ bins, $R(\theta_i,\,\Omega_i)$ is the actual density residual in the $i$-th bin and $\lambda$ is a parameter that characterizes the uncertainties in the model. We fit the $\Omega-\theta$ space residuals over the range $\Omega_{\rm min}=45\,{\rm km/s/kpc}$ and $\Omega_{\rm max}=\Omega_0$ and $-\pi<\theta\leq\pi$. The Fourier coefficients $A_n$ are given by
\begin{equation}
    A_n=\frac{\int_{\Omega_{\rm min}}^{\Omega_{\rm max}}d\Omega
    \int_{-\pi}^\pi R(\theta,\Omega)\cos{\left[n\left (\theta_i-\Omega_i t-\theta_0\right )\right]}d\theta~}{\pi(\Omega_{\rm max}-\Omega_{\rm min})}
\end{equation}
An \textsc{emcee} calculation of the likelihood function yields the best-fit parameters $t = 556\pm 3\,{\rm Myr}$ and $\theta_0 = 2.2\pm 0.2\,{\rm rad}$ for $n_{\rm max} = 4$, though the results are virtually the same for $n_{\rm max}=1,\,2,$ or $3$. Recall that for the perturbed mock data, the true values are $t=500\,{\rm Myr}$ and $\theta_0 = -\pi/2\,{\rm rad}$. Thus, we recover $t$ to about 10\%, but don't recover $\theta_0$. Given that $\Omega\sim 55\,{\rm km/s/kpc}$ for our dataset, this is not surprising as it only takes an error of $\Delta t\simeq50\,{\rm Myr}$, that is, a fractional error of $\simeq10\%$ for $t$, to scramble the value of $\theta_0$ by $\pi$.
In Figure \ref{fig:mock_AAV_contour}, we show the parameter contour plot of the residual fitting in $\Omega(E_z)-\theta$ space. One can see that $t$ and $\theta_0$ are clearly degenerate.

\section{Giant GDR2 sample}\label{GDR2Reduction}

In this section, we describe the steps needed to apply our fitting algorithm to GDR2 data. These include sample selection, a completeness check, and a method for handling measurement uncertainties. 

\subsection{Star catalog and quality cuts}\label{SampleAndCuts}

Historically, the choice of stellar populations to study the local vertical force has been guided by the availability of accurate position and velocity measurements as well as the notion that old populations are well-mixed since they will have made many oscillations through the Galactic plane. Common choices include main sequence stars, such as $F$, $G$, and $K$ Dwarfs \citep{hill1979, bahcall1984a, Kuijken1989b, Zhang2013, Xia2016, Guo2020}, and $K$ Giants \citep{bahcall1984b, Kuijken1989c, holmberg2004}. More recently, \citet{Bienayme2014,hagen2018} and \citet{salomon2020} considered red clump stars which have the advantage that their distances can be accurately determined from photometry since they are good standard candles \citep{Groenewegen2008,Girardi2016,Hawkins2017,Ruiz2018}.

The Gaia mission aims to determine the positions and velocities for $\sim1.2$ billion stars. Already, the radial velocity sample from GDR2 provides complete measurements for the phase space components for $\sim7$ million stars. 
In this study, we draw our sample of giants from the {\it gaiaRVdelpeqspdelsp43} catalog constructed by \citet{sch_prlx}\footnote{See https://zenodo.org/record/2557803 for their data}. The authors show that GDR2 parallaxes are systematically biased and propose corrected parallaxes given by $\varpi_S= \varpi_G+\sqrt{(0.043{\rm mas})^2+{\error_\varpi}^2}$ where $\varpi_G$ and $\error_\varpi$ are GDR2 parallaxes and their uncertainties. We use their distance expectation value, {\it E\_dist}, as a distance estimate. We note that differences between {\it E\_dist} and $1/\varpi_S$ are typically less than 1\%.
We take distance uncertainties to be
\begin{equation}
\varepsilon_r=\sqrt{\langle r^2\rangle-\langle r\rangle^2}
\end{equation}
where $\langle r\rangle$ is given by {\it E\_dist} and $\langle r^2\rangle$ is given by the second moment of the distance probability distribution, {\it distm2}.
We calculate the $z-v_z$ coordinates using the \texttt{astropy.coordinates} Python package\footnote{See https://docs.astropy.org/en/stable/coordinates/index.html for documentation.} where we assume the Sun's distance to the Galactic center as $R_0=\,8.3\,{\rm kpc}$ \citep{Gillessen2009}, the Sun's vertical displacement from the mid-plane as $z_\odot=20.3\,{\rm pc}$ \citep{bennett2019}, and the Sun's vertical motion with respect to the local standard of rest as $v_{z,\odot} = 7.24\,{\rm km/s}$ \citep{schonrich2010}.

We implement the following quality cuts as recommended by \citet{sch_prlx} to ensure better parallax precision:
\begin{itemize}
\item $3<G<14.5$, $G_{RP}>0$, $G_{BP}>0$ where $G_{BP}$, $G$ and $G_{RP}$ are
  apparent magnitudes in Gaia's three broad colour bands. 
  We choose a limit of $G=3$ at the bright end since the whole GDR2 is incomplete for $G<3$ \citep{bennett2019}.
\item $\error_{v_{\rm rad}}<10\,{\rm km/s}$
where $\error_{v_{\rm rad}}$ is the uncertainty in the radial velocity
\item $\error_\varpi<0.1\,{\rm mas}$ and $\varpi/\error_\varpi > 5$ where $\error_\varpi$ is the uncertainties of the parallax measurement.
The second of these cuts is in accord with other papers that have analyzed {\it Gaia} DR2 data \citep{antoja2018,bennett2019,Guo2020,Li2020}.
\item visibility period $n_{\rm vis}>5$
\item $1.172<\verb!bp_rp_excess_factor!<1.3$
\item $d> 80\,{\rm pc}$ where $d$ is the distance from the Sun. This cut reduces systematic distance errors to $<4\%$ \citep{sch_prlx}.
\end{itemize}
In addition to these cuts, we remove stars with Galactocentric speed $\left|{\bf v}\right|>550\,{\rm km/s}$ since these stars have speeds close to or exceeding the escape speed of the Galaxy at the Solar circle \citep{williams2017, monari2018,marchetti2019}. We also exclude stars identified by \citet{vradcontam} as potentially having large radial velocity errors due to contamination of their spectra by a star in close alignment \footnote{See https://arxiv.org/src/1901.10460v1/anc/ for a catalog of these stars}. 
We exclude stars within $15^\circ$ of the Galactic mid-plane (i.e. $|b|<15^\circ$) to avoid losing stars due to obscuration \citep{Katz2019}.

\subsection{Sample volume and CMD region}\label{SampleVolCMD}

In their discovery paper on phase spirals, \citet{antoja2018} selected stars from an annular wedge with Galactocentric radius $8.24\,{\rm kpc} < R < 8.44\,{\rm kpc}$ and Galactic azimuthal angle $\varphi$ within $4^\circ$ of the Sun. Thus, their volume has an extent in the azimuthal direction more than five times larger than the extent in the radial direction. They use all stars from the GDR2 radial velocity survey, which has a magnitude limit at the faint end of $G=17$. However, for all stellar populations combined, the GDR2 radial velocity survey is only complete for $4\lesssim G\lesssim 12.5$ \citep{Katz2019}. Thus, their $z-v_z$ number count map has a $z$-dependent selection function which isn't accounted for.

For our analysis, we carve out a region with the same shape (arc in Galactocentric coordinates) as \citet{antoja2018}, but extend the range in radius to $7.8\,{\rm kpc} < R < 8.8 \,{\rm kpc}$ (i.e. $|R-R_0|<0.5\,{\rm kpc}$) while keeping the same range in $\varphi$. In this way, the extent of of the sample volume in the azimuthal and radial directions ars similar.
As for $z$ and $v_z$, we use stars with $80\,{\rm pc}<|z-z_\odot|<1.48\,{\rm kpc}$ (instead of $|z|<1.48\,{\rm kpc}$ for our mock data) and $|v_z|<80\,{\rm km/s}$ for our fitting. We modify the $z$ selection criteria for two reasons: (1) the sample volume is Sun-centered, and (2) there is little volume left to analyze within $|z-z_\odot|<80\,{\rm pc}$ compared to the whole sample volume since we have removed the region within $80\,{\rm pc}$ from the Sun and also within $15^\circ$ of the mid-plane.

By considering a sample volume that is larger than that of \citet{antoja2018}, we are able to choose stars from a region of the CMD that guarantees completeness for $3<G<14.5$ while maintaining an adequate sample size. Since giant stars are more likely to have available RVS due to their intrinsic brightness, we select a sample of giants with $G_{BP}-G_{RP}>1$ and $M_G<2$ on the CMD. 
For our sample volume, the distance to the Sun ranges between $r_{\rm min}=80\,{\rm pc}$ and $r_{\rm max}=1.67\,{\rm kpc}$. Given our apparent magnitude cut $3<G<14.5$, we apply an absolute magnitude cut of
\begin{equation}
\begin{split}
    3-5{\rm log_{10}}\frac{r_{\rm min}}{10\,{\rm pc}}&=-1.52<M_G<\\
    14.5-5{\rm log_{10}}\frac{r_{\rm max}}{10\,{\rm pc}}&=3.39
\end{split}
\end{equation}
to avoid Malmquist Bias. Combined with the cut $M_G<2$ for giants, we arrive at our absolute magnitude cut of $-1.52<M_G<2$, which concludes all cuts we apply on our data. Our final sample comprises 108,852 stars.

\begin{figure}
	\includegraphics[width=\columnwidth]{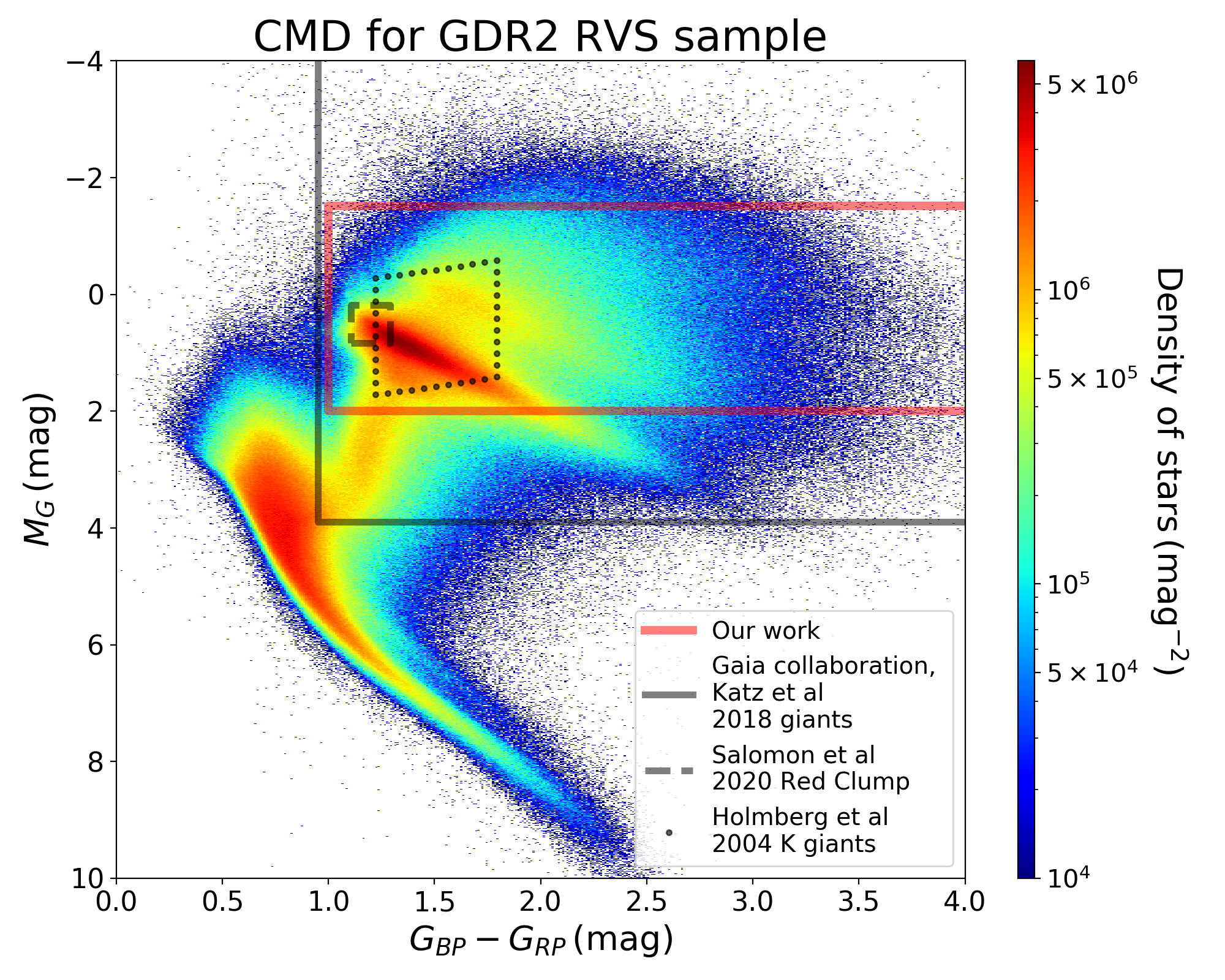}
    \caption{The Color-Magnitude Diagram of RVS-available GDR2 stars with the selection criteria from our work and some other studies as discussed in the text.}
    \label{fig:GDR2-RVS_CMD}
\end{figure}

In Figure \ref{fig:GDR2-RVS_CMD}, we show the region of the CMD used in this work as well as other studies of stellar kinematics in the vicinity of the Sun.
For example, \citet{gaiadr2_vrad} constructed a giant star catalog from GDR2 that was selected by having $G$-band absolute magnitudes of $M_G < 3.9$ and intrinsic colors of $G_{BP}-G_{RP} > 0.95$. This sample of over three million sources was then used to produce mean velocity and velocity dispersion maps in a volume extending to $\sim 4$ kpc from the Sun. Since the authors are computing moments of the DF, completeness issues are second order effects. 
\citet{holmberg2004} consider a sample of K giants from the {\it Hipparcos} catalog selected to be stars with $0 < M_V < 2$ and $1 < B-V < 1.5$. We convert $M_V$ and $B-V$ to $M_G$ and $G_{BP}-G_{RP}$ according to Sect~5.3.7 of the GDR2 Documentation\footnote{See https://gea.esac.esa.int/archive/documentation/GDR2/. We use Johnson-Cousins B-V values as a proxy for Hipparcos B-V since Figure \ref{fig:GDR2-RVS_CMD} is only for illustrative purposes.} in order to show the corresponding box in our Figure \ref{fig:GDR2-RVS_CMD}.
Finally, we present the cut in \citet{salomon2020} as $1.107<G_{BP}-G_{RP}<1.291$ and $0.185<M_G<0.83$, which they use in their Jeans equation analysis of Red Clump stars. 

\subsection{Vertical number density profile and sample completeness}\label{CompCheck}

\begin{figure}
	\includegraphics[width=\columnwidth]{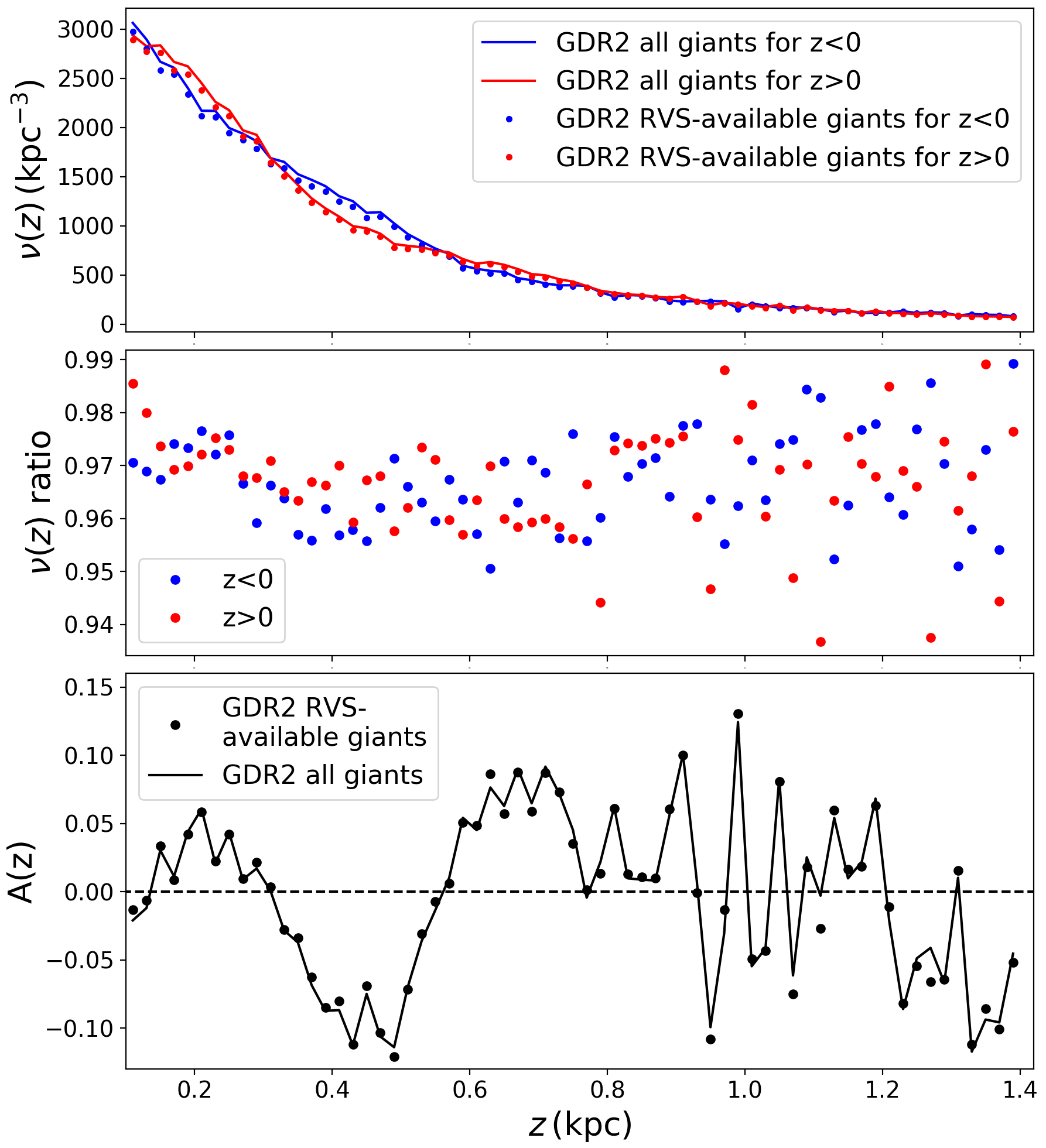}
    \caption{The stars' $z$ distribution profile with detailed descriptions in the text. Note that $A(z)$ is a measure of vertical number density asymmetry that reflects the fractional difference of North-South number density.}
    \label{fig:completeness}
\end{figure}

We examine the distribution of stars as a function of $z$ in Figure \ref{fig:completeness}.
In the upper panel, we show the volume number density profile $\nu(z)\equiv n(z)/{\cal G}(z)$ 
for both our sample and a sample that includes stars that do not necessarily have radial velocity measurements. 
For stars that are not in the \textit{gaiaRVdelpeqspdelsp43catalog} catalogue, we estimate distances following the procedure outlined in \citet{sch_prlx}, that is, $d = 1/\varpi_S$ where $\varpi_S = \varpi_G+\sqrt{(0.043\,{\rm mas})^2+\varepsilon_\varpi^2}$. In addition, we highlight differences between the number counts north and south of the mid-plane by
plotting $\nu(z)$ separately for $z>0$ and $z<0$. 
The middle panel shows the ratio of $\nu(z)$ from the radial velocity sample and the full sample. The fact that the ratio is close to unity implies that the radial velocity sample is as complete as the full GDR2 survey. That is, for our giant sample, the restriction to stars with radial velocity measurements doesn’t introduce any new selection effects.
In the lower panel, we show our results for the North-South asymmetry, $A(z)=\left[\nu(z)-\nu(-z)\right]/\left[\nu(z)+\nu(-z)\right]$, which is consistent with what has been found in \citet{widrow2012, bennett2019,salomon2020}. This asymmetry has been interpreted as evidence for disequilibrium in the stellar disk.

\subsection{Measurement uncertainties}

Statistical uncertainties in stellar distances, proper motions and radial velocities imply uncertainties in $z$ and $v_z$. Thus, the $z-v_z$ bin assigned to a given star is also uncertain. 
In addition, the true position of a star can lie outside the sample volume when its measured position lies inside it. By calculating the uncertainties in $z$ and $v_z$ (see formulae in \citet{johnson1987}) we estimate that $\simeq 15\%$ of stars have a true bin different from on the one implied by the measured kinematics.

To account for these uncertainties, we use a bootstrap method to generate 100 datasets where astrometric quantities are sampled from measured quantities under the assumption that the uncertainties are Gaussian. We then convert each dataset to $x,\,y,\,z$ and $v_z$ using the \texttt{astropy.coordinates} Python package and conduct data selection and fitting as previously discussed. Finally, the MCMC chains from each of the datasets are combined to yield a PDF for the model parameters.

\section{Results}\label{GDR2Results}

In Table \ref{tab:result}, we present best-fit values and $1\sigma$ errors for all parameters as well as $\psi(z)$ and $F_z$ at $z=0.5\,{\rm kpc},\,1.0\,{\rm kpc}$ and $1.5\,{\rm kpc}$.  
\begin{table}
    \centering
    \renewcommand{\arraystretch}{1.4}
    \begin{tabular}{cl}
    Quantity & Best-fit value and $1\sigma$ error \\
    \hline
    $\omega_1$ & $52.8_{-1.3}^{+1.4}\,{\rm km/s/kpc}$ \\
    $D$ & $0.36_{-0.06}^{+0.07}\,{\rm kpc}$ \\
    $\omega_2$ & $31.4_{-3.4}^{+2.5}\,{\rm km/s/kpc}$ \\
    $\sigma_z$ & $12.8\pm0.1\,\,{\rm km/s}$ \\
    $\ln\,\alpha$ & $0.97\pm0.01$ \\
    $\psi$(0.5 kpc) & $376.1\pm4.6\,{\rm (km/s)^2}$ \\
    $\psi$(1.0 kpc) & $1187\pm9\,{\rm (km/s)^2}$ \\
    $\psi$(1.5 kpc) & $2276_{-40}^{+39}\,{\rm (km/s)^2}$ \\
    $\fz(0.5\,{\rm kpc})$ & $1294\pm16\,{\rm (km/s)^2/kpc}$ \\
    $\fz(1.0\,{\rm kpc})$ & $1914\pm35\,{\rm (km/s)^2/kpc}$ \\
    $\fz(1.5\,{\rm kpc})$ & $2438_{-114}^{+102}\,{\rm (km/s)^2/kpc}$\\
    $\Sigma(0.5\,{\rm kpc})$ & $44.6\pm0.8\,{\rm M_\odot pc^{-2}}$\\
    $\Sigma(1.0\,{\rm kpc})$ & $64.3\pm1.7\,{\rm M_\odot pc^{-2}}$\\
    $\Sigma(1.5\,{\rm kpc})$ & $80.4_{-4.5}^{+4.1}\,{\rm M_\odot pc^{-2}}$
    \end{tabular}  
    \caption{Best-fit values and $1\sigma$ errors of all parameters as well as $\psi(z)$, $F_z$ and $\Sigma(z)$ at $z=0.5\,{\rm kpc},\,1.0\,{\rm kpc}$ and $1.5\,{\rm kpc}$.}
    \label{tab:result}
\end{table}
We also present $\Sigma(z)$ at these heights, which are derived from Equation \ref{eq:Sigmaz}. To do so, we assume $A=15.45\pm0.34\,{\rm km/s/kpc}$ and $B=-12.27\pm0.40\,{\rm km/s/kpc}$, which are averages of recent measurements by \citet{Bovy2017, Vityazev2017, Bobylev2018, Nouh2020, Krisanova2020}. 

\begin{figure}
	\includegraphics[width=\columnwidth]{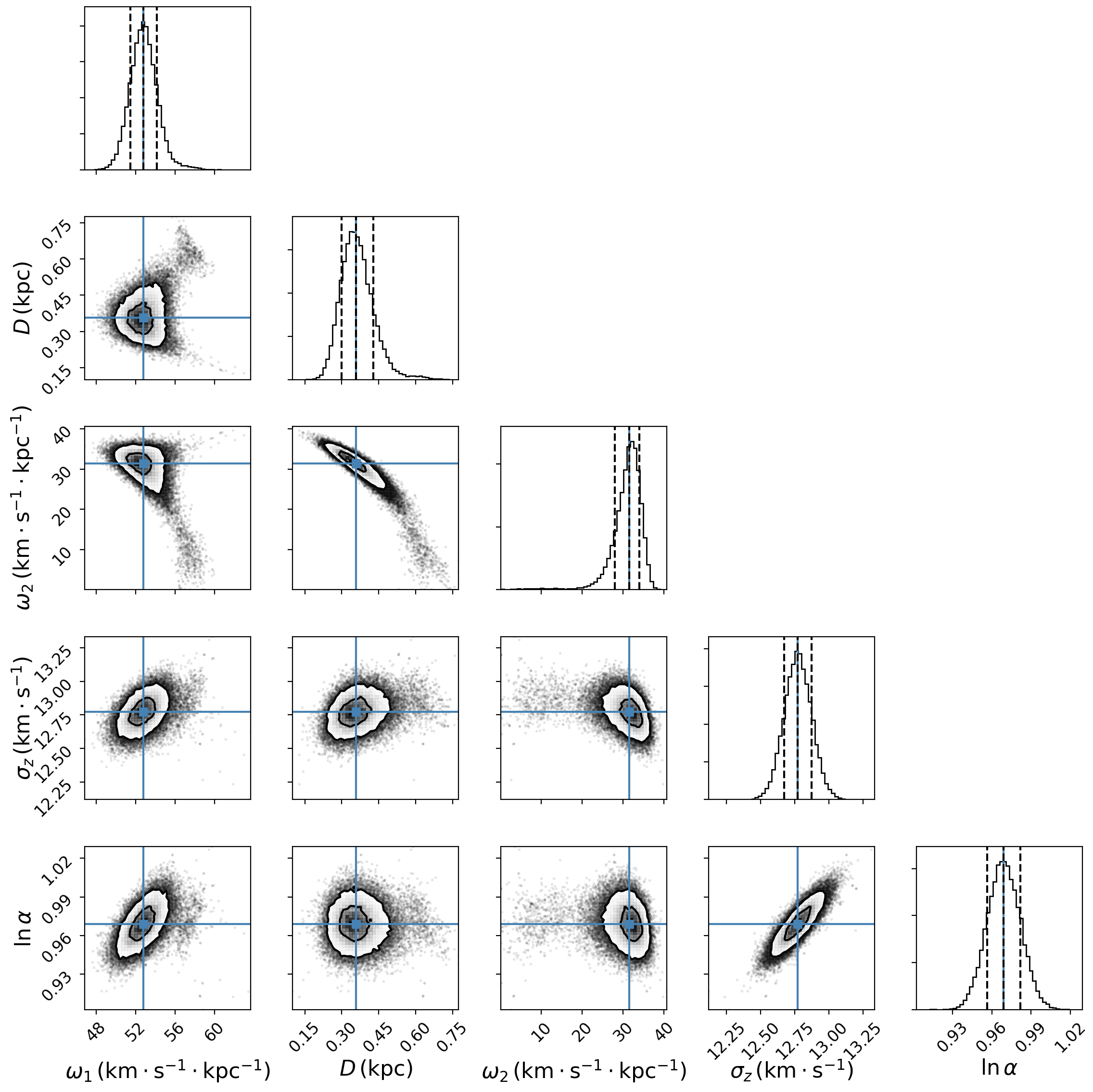}
    \caption{Parameter contours of our fitting. The blue lines indicate best-fit values presented in Table~\ref{tab:result}.}
    \label{fig:contour}
\end{figure}

In Figure \ref{fig:contour}, we show one- and two-dimensional projections of the likelihood function via the so-called corner plot \citep{FM2016}.
We see that the peak of the likelihood function is well inside the region defined by the prior probabilities for the parameters given in Table \ref{tab:prior}. The likelihood function appears to be reasonably well behaved, though there is a tail of outliers, which will be discussed in detail in Section \ref{ParaDegen}. Along the tail, $D$ is large and therefore the two terms in Equation \ref{eq:KGPotential} can both be regarded as quadratic. Hence, $\omega_1$ and $\omega_2$ are approximately degenerate and therefore anti-correlated. 
In addition, a positive correlation exists between $\alpha$ and $\sigma_z$. To see this, consider Equation \ref{eq:RLDFdispersion} in the limit $z=0$. We then have $\langle {v_z}^2\rangle_0 = {\sigma_z}^2\left(1-\frac{3}{2\alpha}\right)^{-1}$. Thus, for fixed mid-plane dispersion, an increase in $\alpha$ requires an increase in $\sigma_z$.

\begin{figure}
	\includegraphics[width=\columnwidth]{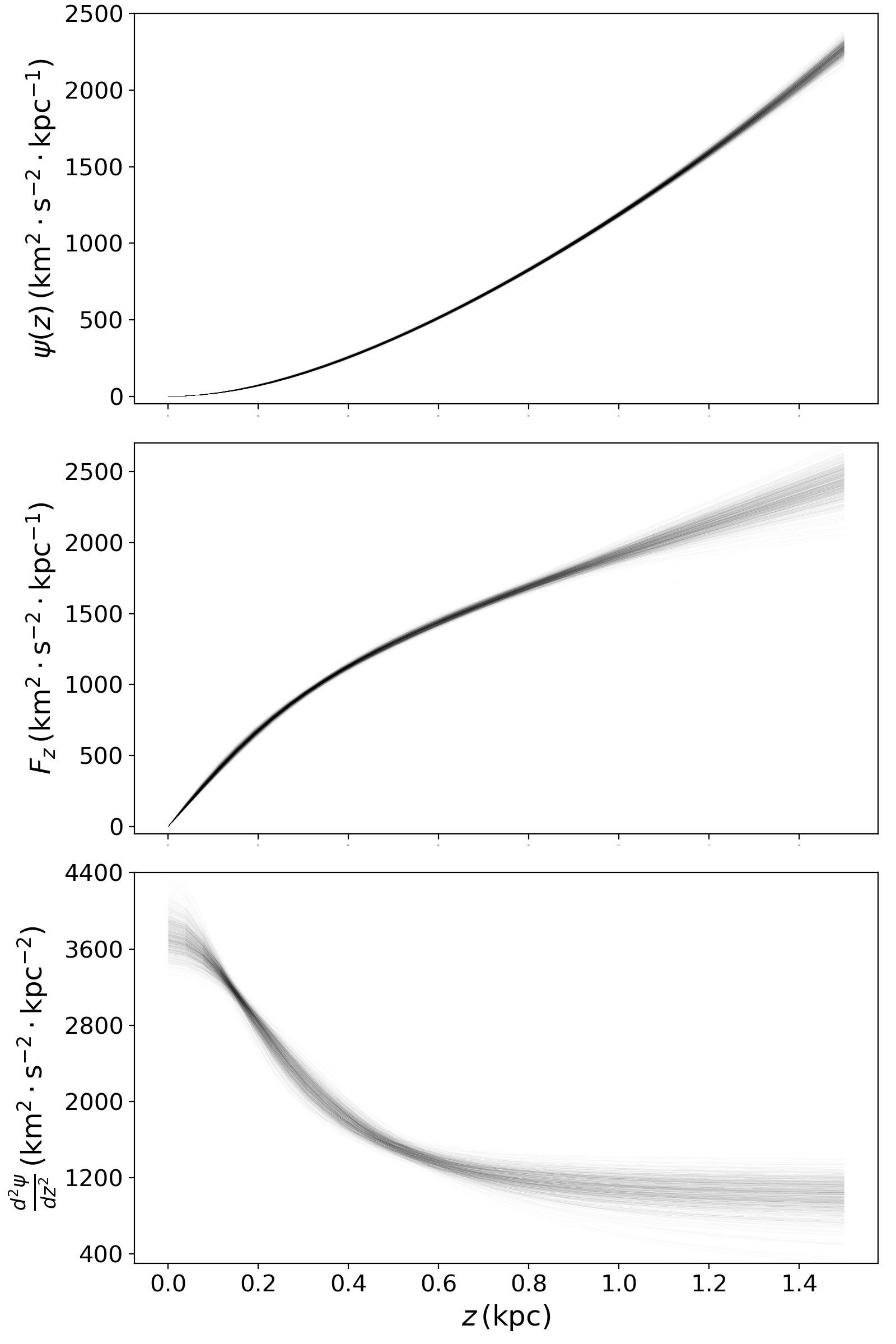}
    \caption{The profiles of $\psi(z)$, $F_z$ and $d^2\psi/dz^2$ derived from the fitting for our GDR2 data. We show 1000 random samples selected from the MCMC sampling for each panel.}
    \label{fig:GaiaResult}
\end{figure}

\begin{figure}
	\includegraphics[width=\columnwidth]{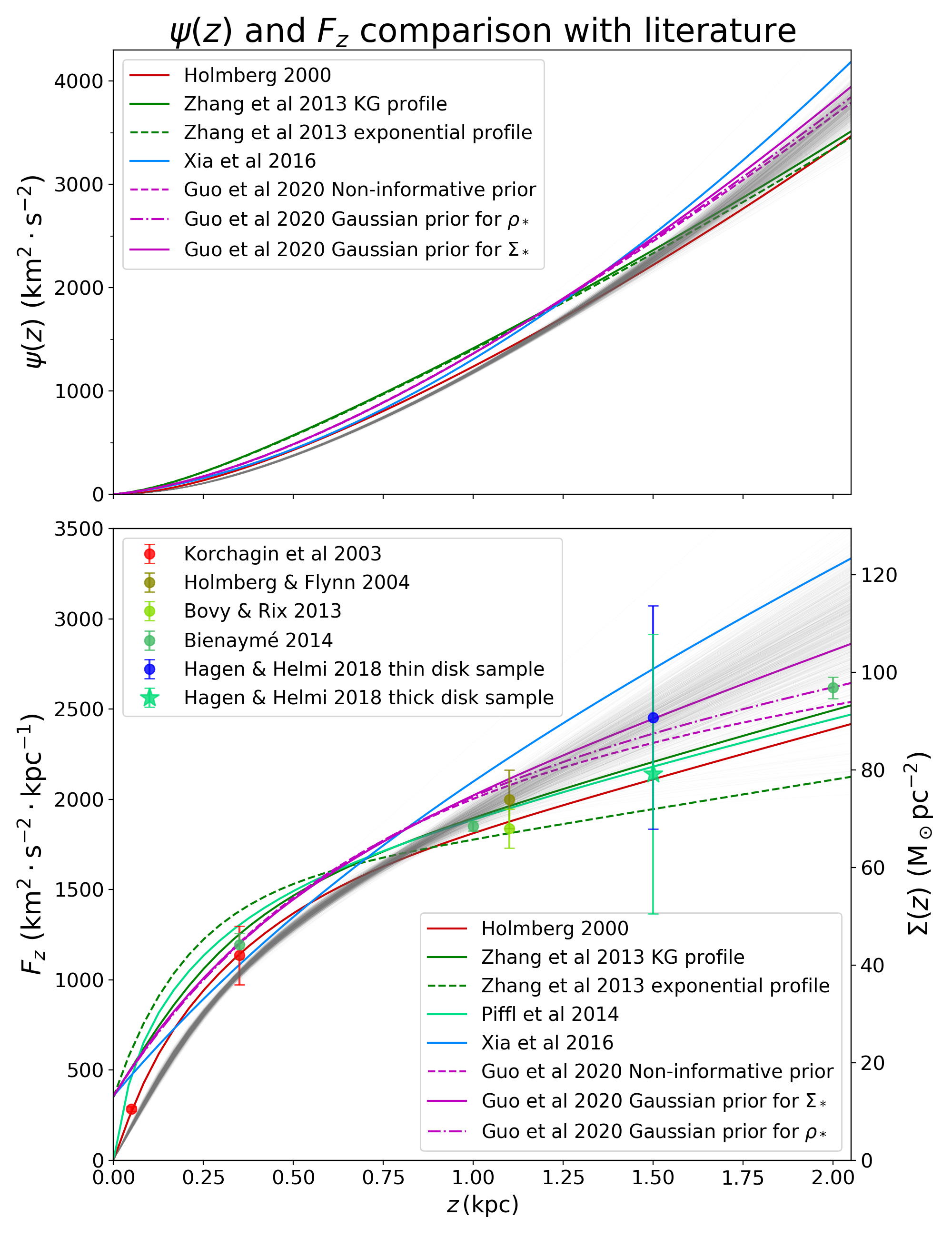}
    \caption{Comparison of our vertical potential and force with literature values. The grey lines are derived from 1000 random samples in our parameter sampling while literature values from \citet{holmberg2000,holmberg2004,bovy2013,Zhang2013,Bienayme2014,piffl2014, Xia2016,hagen2018} and \citet{Guo2020} are indicated with other legends. For the force panel, we also label $\Sigma(z)\simeq\frac{F_z}{2\pi G}$ values as an indication of surface density. Note that we don't include the radial effects here as it is a widely applied assumption to ignore such effect in the literature. The non-zero force at $z=0$ for some of the curves results from a razor-thin gas with surface density is $\Sigma_g\simeq13\,{\rm M_\odot pc^{-2}}$ that was included in those models}.
    \label{fig:PotFCompare}
\end{figure}

In Figure \ref{fig:GaiaResult}, we show profiles of $\psi$, $F_z$, and $d^2\psi/dz^2$, which are calculated using 1000 samples of the model parameters from the MCMC chain.
In Figure \ref{fig:PotFCompare}, we compare our $\psi(z)$ and $F_z$ profiles with those obtained by
\citet{holmberg2000,holmberg2004,bovy2013,Zhang2013,Bienayme2014,piffl2014, Xia2016,hagen2018} and \citet{Guo2020}.
Our results for $\psi(z)$ are similar to those from the literature. As expected, differentiation amplifies the differences as can be seen in the lower panel of the Figure where we show $F_z(z)$. Our estimate for the force is generally lower than the literature values for $|z|\lesssim 700\,{\rm pc}$ but is consistent with the literature values for $z\simeq 800 {\rm pc}$, which is also where the scatter in the models is at a minimum. At larger values of $z$ the scatter increases due to the lack of data. 

Based on these results we expect that our estimate for the matter density near the mid-plane will be below the values quoted in the literature. 
Indeed, we obtain $\rho_0=0.067_{-0.003}^{+0.004}\,{\rm M_\odot pc^{-3}}$ from equation \ref{eq:rho0} with a correction of $-0.003\,{\rm M_\odot pc^{-3}}$ from the radial term $2(B^2-A^2)$. Our value is roughly 30\% lower than typical literature values which cluster around $0.09\,{\rm M_\odot pc^{-3}}$ (\citealt{Kuijken1989c,holmberg2000,Bienayme2014,McKee2015}).

Recently, \citet{chakrabarti2021} estimated the vertical force as a function of $z$ using pulsar timing measurements. In particular, they
found $\log_{10}(\alpha_1/{\rm Gyr}^{-2}) = 3.69^{+0.13}_{-0.10}$ where
$\alpha_1$ is the slope of the vertical force as a function of $z$ near the mid-plane and equal to our $\omega_1^2 + \omega_2^2$ in our parameterization of the potential. Their result is consistent with our value of $\log_{10}((\omega_1^2 + \omega_2^2)/{\rm Gyr}^{-2})=3.59\pm 0.02$. However, we caution that our low value for $\omega_1^2 + \omega_2^2$ may be biased by the inclusion of stars with high in-plane velocity dispersion for which the 1D approximation is suspect. (See discussion below.)

\begin{figure}
	\includegraphics[width=\columnwidth]{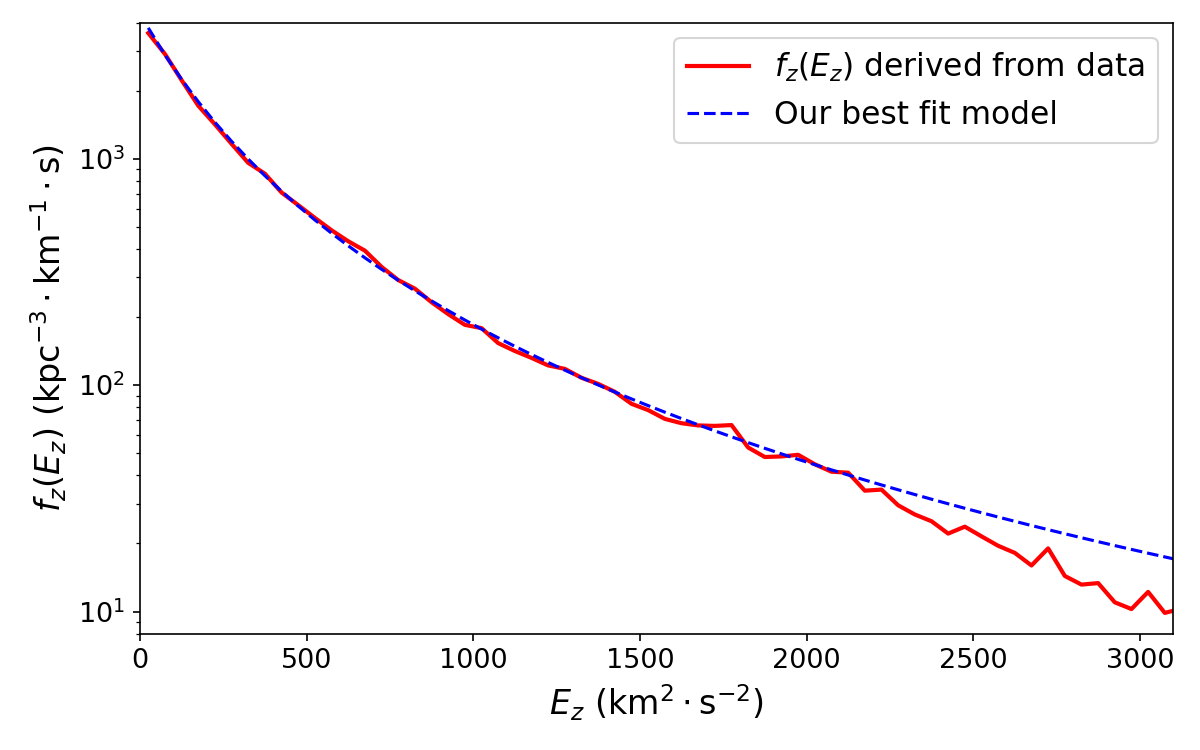}
    \caption{Vertical distribution function $f_z\left (E_z\right )$ for the data and model. The solid red curve is the DF for the stars in our sample where the vertical energy is calculated using best-fit parameters for the potential from Table \ref{tab:result}. The dotted blue curve shows our RLDF model (Equation \ref{eq:RLDF}) with $\alpha = 2.64$ and $\sigma_z = 12.8\,{\rm km\,s^{-1}}$.}
    \label{fig:fofe}
\end{figure}

In Figure \ref{fig:fofe} we compare $f_z$ from our model with that derived from the data. The vertical energy is calculated by assuming that $\psi(z)$ is given by Equation \ref{eq:KGPotential} with the parameters $\omega_1$, $\omega_2$ and $D$ taken from Table \ref{tab:result}. For the model, $f_z$ is given by Equation \ref{eq:RLDF} with $\sigma_z = 12.8\,{\rm km/s}$ and $\alpha = \exp\, (0.97)=2.64$. The data shows the well-known trend that the velocity dispersion smoothly increases with increasing $E_z$. Our RLDF model provides an excellent fit to the data for $E_z \lesssim 2100 {\rm (km/s)^2}\simeq1/2\times \left(65\,{\rm km/s}\right)^2$.

\begin{figure}
	\includegraphics[width=\columnwidth]{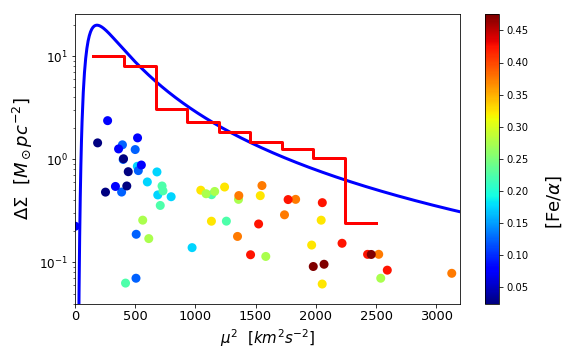}
    \caption{Vertical temperature distribution in the Solar Neighbourhood from an analysis of SDSS/SEGUE G-dwarf data by \citet{bovy2012b} and from our analysis of GDR2 giants. The points and histogram are the same as in Figure 8 in \citet{bovy2012b}. The points show the surface density for mono-abundance populations as a function of their vertical velocity dispersion squared, ${\mu_z}^2$. Colors indicate the $[Fe/\alpha]$ abundance. The histogram gives the contribution to the surface density, $\Delta \Sigma$ for bins of width $\Delta{\mu_z}^2 = 260\,{\rm km^2\,s^{-2}}$. 
    The solid blue curve is our model prediction for $\Delta\Sigma
    =\left (\Delta \mu_z^2/2\mu_z\right )\left (d\Sigma/d\mu_z\right )$
    where $d\Sigma/d\mu_z$ is given by Equation \ref{eq:frac_Sigma} with $\sigma_z = 12.8\,{\rm km\,s^{-1}}$ and $\alpha = 2.64$. We normalize our curve so that the area under our curve equals the area under the histogram.}
    \label{fig:sigsig}
\end{figure}

In Figure \ref{fig:sigsig} we show our prediction for the vertical temperature distribution in terms of $\frac{d\Sigma}{d{\mu_z}^2}=\frac{d\Sigma}{2\mu_z d\mu_z}$. Recall that $\mu_z$ corresponds to the vertical velocity dispersion of the isothermal constituents that make up the RLDF though Equation \ref{eq:RLDF_decompose}. 
Apart from a sharp rise with $\mu_z$ near $\mu_z = 0$, $d\Sigma/d\mu_z^2$ is a decreasing function of $\mu_z$. \citet{bovy2012b} approximate this function, which they call the vertical temperature distribution, for a sample of G-dwarf from SDSS/SEGUE \citep{abazajian2009, yanny2009}. They arrive at their estimate by first dividing the sample into mono-abundance sub-populations as defined by $[\alpha/{\rm Fe}]$ and $[{\rm Fe}/H]$. These sub-populations are found to have a velocity dispersion that is approximately constant in $z$, which implies that the sub-populations are approximately isothermal. 
A scatter plot of the surface density for each sub-population as a function of dispersion and $[\alpha/{\rm Fe}]$ is shown in their Figure 8 and reproduced here in Figure \ref{fig:sigsig}.  Also shown is the histogram derived by binning the sub-populations in ${\mu_z}^2$. We see that the distributions indicated by our curve and their histogram are qualitatively similar though they are derived for different populations of stars.

\begin{figure}
	\includegraphics[width=\columnwidth]{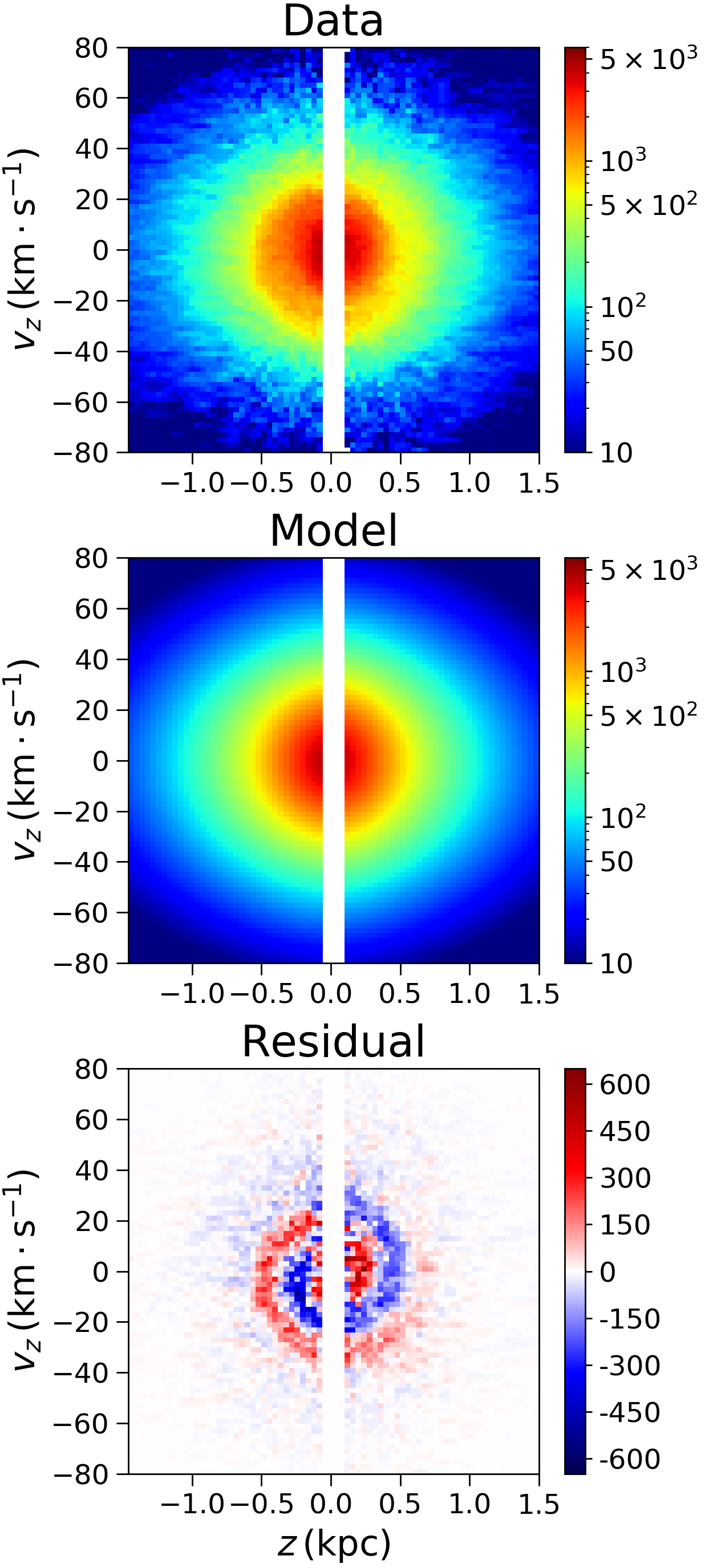}
    \caption{Density in $z-v_z$ plane of our GDR2 sample (upper panel) and our best-fit model (middle panel) along with the fitting residual (lower panel) in unit of ${\rm kpc^{-3}\cdot{km/s}^{-1}}$. The overdensity in the residual panel forms a clear spiral pattern.}
    \label{fig:GaiaDF}
\end{figure}

\begin{figure}
	\includegraphics[width=\columnwidth]{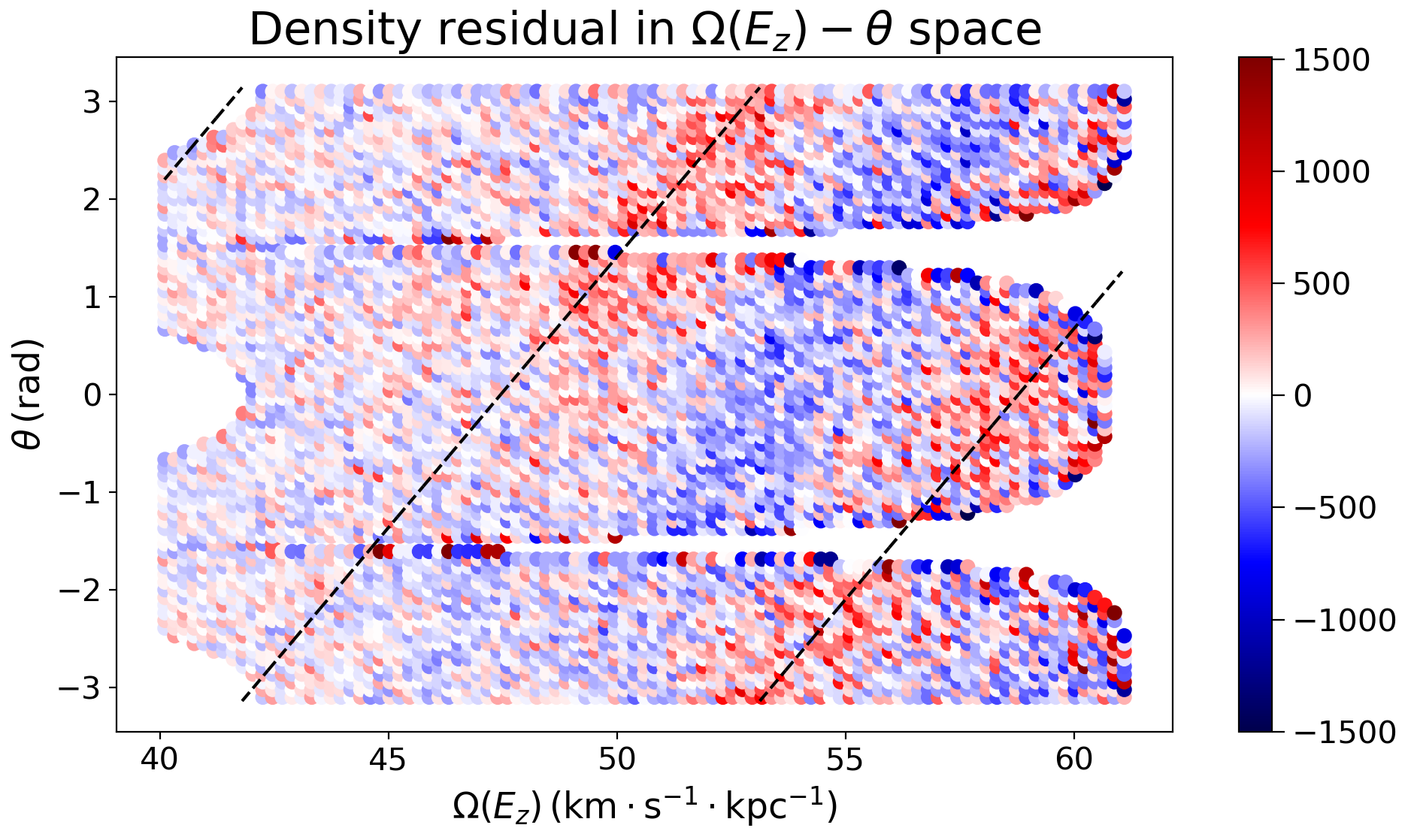}
    \caption{The residual of phase-space density plotted in $\Omega-\theta$ space. Parallel straight lines corresponding to $t=543\,{\rm Myr}$ and $\theta_0=-1.2\,$rad are overplotted in black. The unit of the residual is ${\rm kpc^{-1}\cdot s\cdot km^{-1}\cdot rad^{-1}}$.}
    \label{fig:spiralAAV}
\end{figure}

In Figure \ref{fig:GaiaDF} we plot the number densities in the $z-v_z$ plane for data and our best-fit model, along with the residuals defined as ${\rm data}-{\rm model}$. We take the sample geometry into account by dividing number counts by the geometric factor ${\cal G}(z)$. Thus, the quantity presented has dimensions of stars per ${\rm volume\times velocity}$. The residuals in Figure \ref{fig:GaiaDF} show the phase spiral discovered by \citet{antoja2018} whose locations are consistent with what found by \citet{Li2021} at a guiding radius of $8.34\,{\rm kpc}$.
In Figure \ref{fig:spiralAAV} we plot the residuals in $\Omega(E_z)-\theta$ space as we did for the mock data in Figure \ref{fig:mockAF}. The region $|z-z_\odot| < 80\,{\rm pc}$, which is omitted from the sample volume, maps into two bands centered on $\theta = \pm\pi/2$.
A fit using the procedure described in Section \ref{mock_AAV_fitting} yields the parallel straight lines in Figure \ref{fig:mockAF} with $t = 543 \pm 11\,{\rm Myr}$ and $\theta_0 = -1.2 \pm 0.6 \,{\rm rad}$. 
The perturbation age is in agreement with \citet{antoja2018} who estimate that the perturbing event started $\sim 500\,{\rm Myr}$ ago with a likely range of $300\sim900\,{\rm Myr}$.

\section{Discussion}\label{Discussion}

\subsection{Parameter degeneracies}\label{ParaDegen}

\begin{figure}
	\includegraphics[width=\columnwidth]{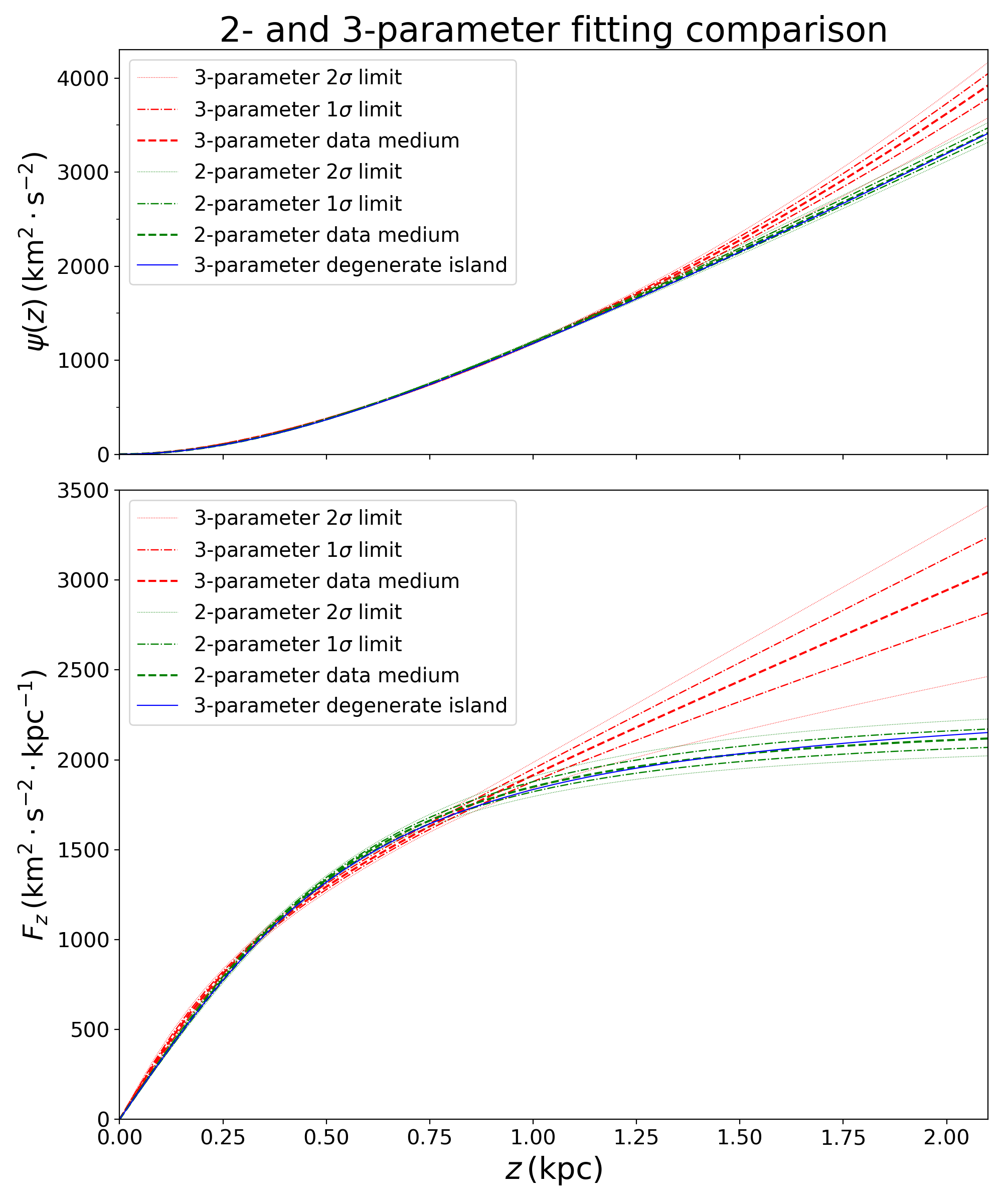}
    \caption{Comparison of $\psi(z)$ and $F_z$ from 2-parameter and 3-parameter potential fitting and the degenerate case of 3-parameter potential.}
    \label{fig:23para}
\end{figure}

As noted in the previous section, the PDF for the model parameters shows a probability island connected to the main peak by a bridge. Since this feature isn't seen in the $\alpha-\sigma_z$ plane, we conclude that it is related to the potential and not the DF. The bridge and island stretch to higher values of $D$. In this region of parameter space, the first term in the potential is approximately quadratic for small and intermediate values of $z$. We therefore expect $\omega_1$ and $\omega_2$ to be anti-correlated for this region and indeed this is what is seen in the $\omega_1-\omega_2$ projection of the PDF. We explore this degeneracy further in Figure \ref{fig:23para} where we plot $\psi$ and $F_z$ for parameters characteristic of the island, namely
$\omega_1=57\,{\rm km/s/kpc}$, $D=0.65\,{\rm kpc}$ and $\omega_2=8\,{\rm km/s/kpc}$. The potential and force are nearly the same as those obtained from our best-fit values from Table \ref{tab:result} for $z \lesssim 900\,{\rm pc}$ but strongly diverge at larger $z$. Also shown are the results from an MCMC analysis where $\omega_2$ is fixed to be zero. As with the ``island" model, the potential and force are nearly the same as those obtained with the three-parameter potential out to $z\sim 900\,{\rm pc}$. 

These results are symptomatic of the well-known fact that it is difficult to disentangle the disc, bulge, and halo contributions from the potential despite many attempts
(see, for example, \citealt{Zhang2013,Xia2016,Sivertsson2018,Guo2020}). As mentioned in Section \ref{AlgoIntro}, \citet{Kuijken1989a, Kuijken1989b} identify the first and second terms in Equation \ref{eq:KGPotential} with the disc and (effective) halo respectively. But it is clear that the potential within the first few scale heights of the disc is adequately fit by a simple two-parameter model. Note that while the potentials are similar, the inferred values for the mid-plane density, which is derived from the second derivative at $z=0$ can vary considerably. We can see this already in our models which show an increase in the scatter of $d^2\psi/dz^2$ as $z\to 0$. In fact, the estimates for ${\omega_1}^2 + {\omega_2}^2$ for our three-parameter and island models differ by $13\%$. The baryon-dark matter degeneracy was illustrated in \citet{Guo2020} who illustrated that their inferred values for the dark matter density in the Solar Neighbourhood were sensitive to the choice of priors for the stellar surface and volume densities (see their Table~2). However, as seen in Figure \ref{fig:PotFCompare}, the potential and force are
relatively insensitive to priors on the stellar component.

\subsection{Separability assumption and mid-plane density}

The analysis presented in this work is based on the 1D approximation wherein the vertical dynamics of Solar Neighbourhood stars is separated from their dynamics in the Galactic plane and that the vertical force is independent of $R$. 
In reality, in-plane and vertical motions are coupled since the full potential $\Psi(R,z)$ is not separable. The effects of a non-separable potential are greatest for stars with large radial and azimuthal velocity dispersion since these stars make the largest excursions from the Solar Neighbourhood. Following \citet{Li2020} we reanalyze our data by first dividing the sample into cold and hot sub-samples. As we will see, the results may shed light on the discrepancy between our inferred value for $\rho_0$ and the values found in the literature.

\begin{figure}
	\includegraphics[width=\columnwidth]{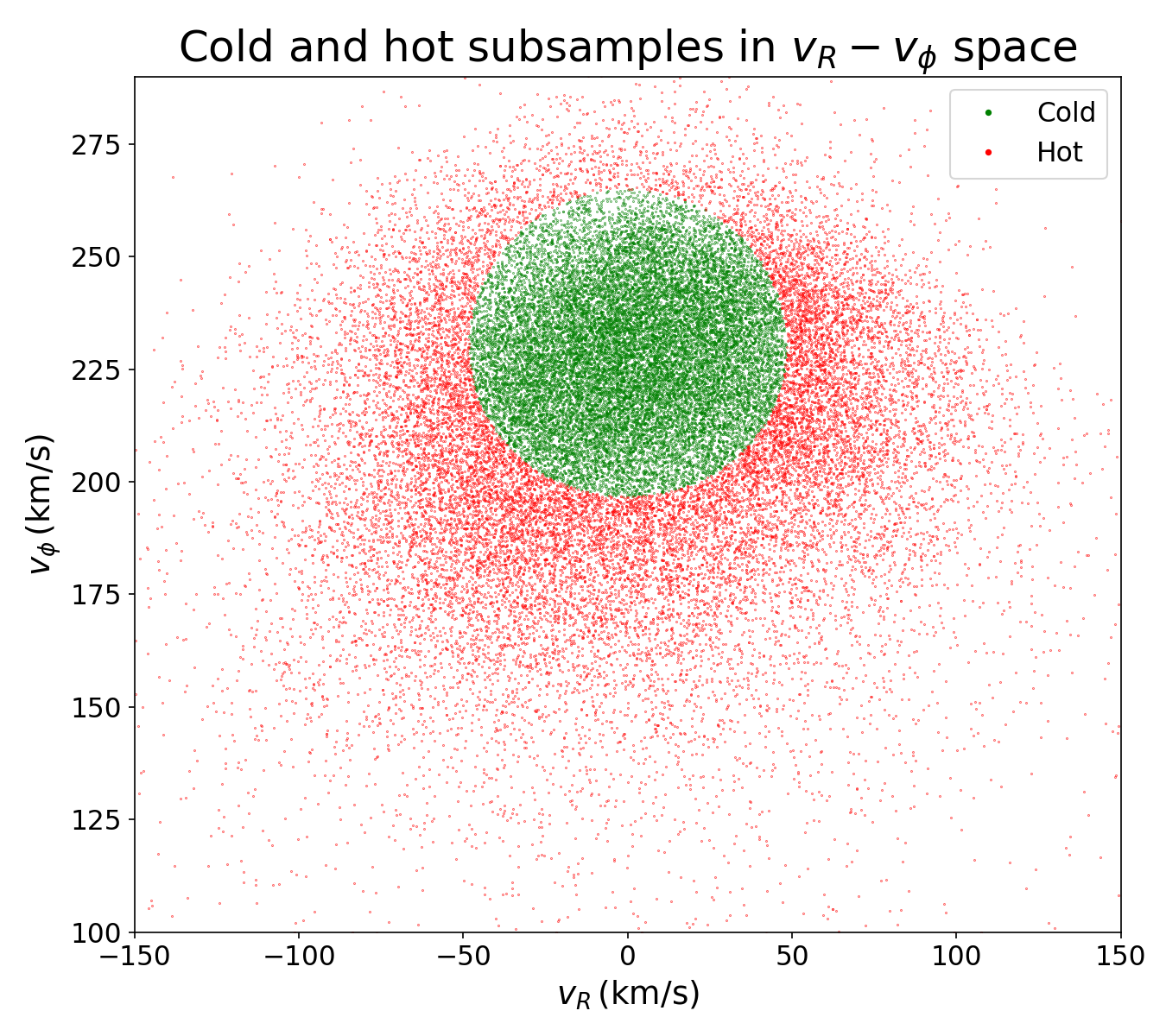}
    \caption{Randomly selected 10 thousand cold (green) and hot (red) stars respectively in the $v_R-v_\phi$ plane.}
    \label{fig:ColdHotVrVphi}
\end{figure}

We define the planar energy of a star as
\begin{equation}
    E_{\rm R,eff}(R,v_R)\equiv\frac{{v_R}^2}{2}+\chi_{\rm eff}(R)
\end{equation}
where $\chi_{\rm eff}(R)\equiv\Psi(R,0)+\frac{{L_z}^2}{2R^2}$ is the effective radial potential and $L_z$ is the vertical angular momentum. We assume that near the Sun, the potential in the mid-plane has the form $\Psi(R,0) = v_c^2\ln{R}$ such that the rotation curve of the Solar Neighborhood is flat with a rotation speed of $v_c$. We set $v_c = 230\,{\rm km\,s}^{-1}$, which is close to the average of  recent measurements by \cite{bovy2009,kop2010,bovy2012c,honma2012,Reid2014,eilers2019}. 
A star with angular momentum $L_z$ and $v_R = 0$ follows a circular orbit with $R=\frac{L_z}{v_c}$.
The epicycle energy, which we define as
\begin{equation}
    E_{\rm epi}\equiv E_{\rm R,eff}(R,v_R)-E_{\rm R,eff}\left(\frac{L_z}{v_c},0\right)
\end{equation}
indicates the extent of the radial motion. We sort stars in our sample by increasing $E_{\rm epi}$ and take first and the second halves to be the cold and hot sub-samples, respectively. The hot sample has larger extent of radial motion and is therefore more affected by the coupling of in-plane and vertical motion. In Figure \ref{fig:ColdHotVrVphi}, we randomly select 10 thousand cold and hot stars respectively from our sample and plot them on the $v_R-v_\phi$ plane. The figure shows a clear elliptical boundary between the two sub samples and is in good agreement with Figure 7 of \citet{Li2020} where the radial action $J_R$ is used to separate hot and cold components.

\begin{figure}
	\includegraphics[width=\columnwidth]{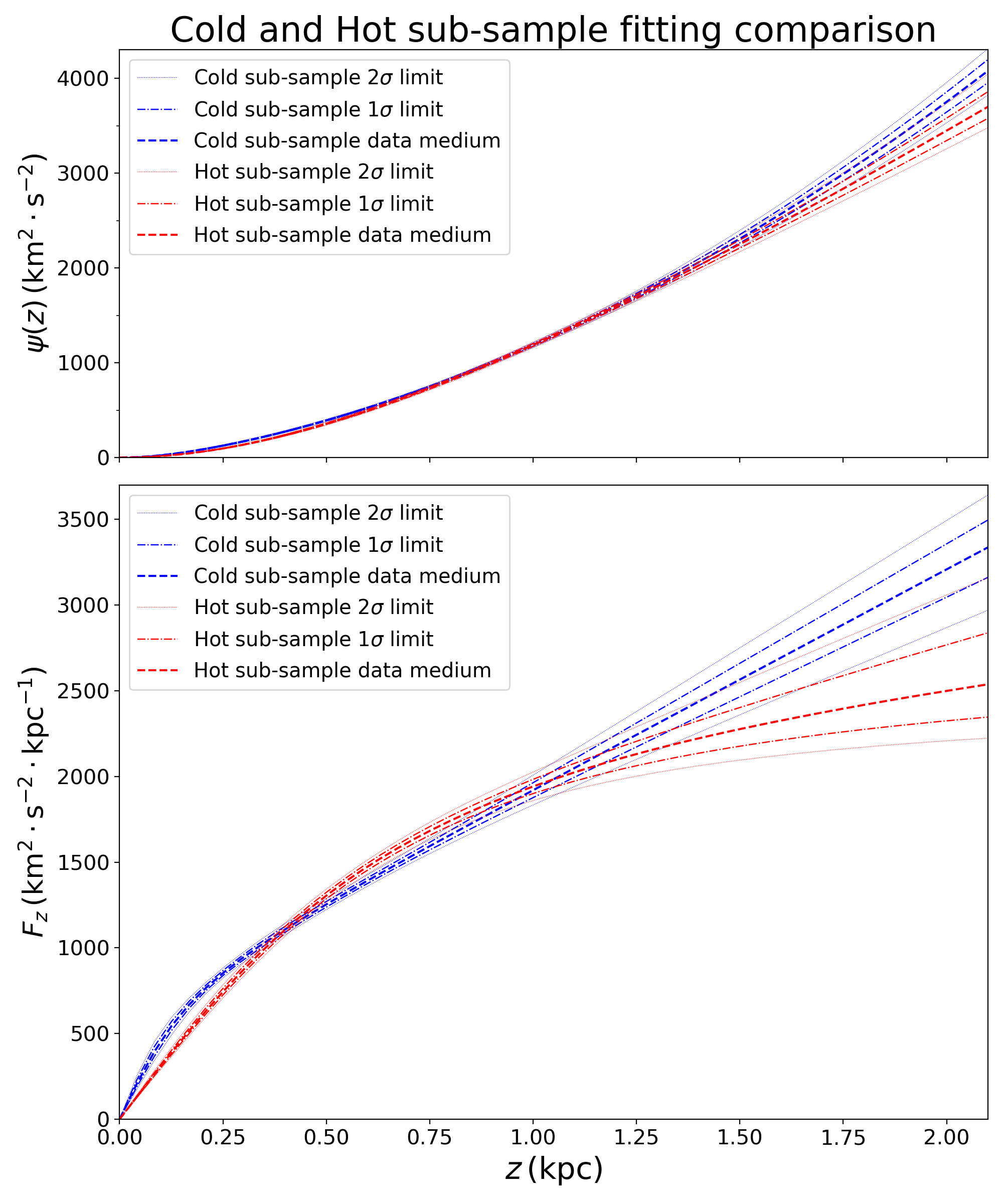}
    \caption{Comparison of $\psi(z)$ and $F_z$ from cold and hot sub-sample fitting.}
    \label{fig:ColdHotCompare}
\end{figure}

Results for our analysis for the two sub-samples are presented in Table \ref{tab:ColdHotPara} as well as $\psi(z)$ and $F_z$ shown as Figure \ref{fig:ColdHotCompare}.  
\begin{table}
    \centering
    \renewcommand{\arraystretch}{1.4}
    \begin{tabular}{ccc}
    Parameter & the cold sub-sample & the hot sub-sample \\
    \hline
    $\omega_1$ & $61.7_{-3.4}^{+4.3}\,{\rm km/s/kpc}$ & $53.3_{-3.0}^{+2.0}\,{\rm km/s/kpc}$  \\
    $D$ & $0.17\pm0.04\,{\rm kpc}$ & $0.70_{-0.11}^{+0.09}\,{\rm kpc}$ \\
    $\omega_2$ & $35.8_{-1.8}^{+1.5}\,{\rm km/s/kpc}$ & $17.1_{-9.4}^{+8.5}\,{\rm km/s/kpc}$ \\
    $\sigma_z$ & $12.0\pm0.1\,\,{\rm km/s}$ & $15.1\pm0.2\,\,{\rm km/s}$ \\
    $\ln\,\alpha$ & $1.23\pm0.02$ & $0.90\pm0.02$ 
    \end{tabular}  
    \caption{Best-fit parameter values and $1\sigma$ errors for the cold and hot sub-sample.}
    \label{tab:ColdHotPara}
\end{table}
The inferred values for the model parameters from the two sub-samples differ significantly. For the mid-plane density,
which is proportional to $\omega_1^2+\omega_2^2$, we find that $\rho_{\rm 0,cold}=0.0942_{-0.009}^{+0.011}\,{\rm M_\odot pc^{-3}}$ and $\rho_{\rm 0,hot}=0.0581_{-0.002}^{+0.002}\,{\rm M_\odot pc^{-3}}$. Our value for $\rho_{\rm 0,cold}$ is in good agreement with those in the literature values while $\rho_{\rm 0,hot}$ is even lower than the value we obtained for the whole sample. We also find that the inferred value of $D$ is much larger for the hot sub-sample than the cold sub-sample. These differences are reflected in the inferred potential and force for the two sub-samples shown in Fig. \ref{fig:ColdHotCompare}. In particular, the force inferred from the hot sub-sample is shallower at the origin and linear out to larger values of $z$ as compared with the force inferred from the cold sub-sample. Finally, the $\alpha$ parameter is smaller for the hot sub-sample. In other words, its DF decreases with energy more slowly than the DF of the cold sub-sample. 

Of course, all stars experience the same potential. The fact that different $\psi(z)$'s are inferred from different sub-samples is an indication that the separability assumptions for the potential and DF are breaking down. In particular, stars in our hot sub-sample make radial excursions from their guiding radii by several kpc; stars from the cold sub-sample stay closer to their guiding radii. Thus, the former experience a vertical potential over a wider range of the Galactic disc. In principle,
these issues could be addressed by moving to a full three-dimensional model for the potential and three-integral model for the DF as in \citet{binney2010, binney2011, piffl2014}.

\section{Conclusions}\label{Conclusion}

In this work, we introduce a method for inferring the local vertical potential and stellar DF from kinematic measurements of stars in the vicinity of the Sun. The method is based on Jeans Theorem under the 1D approximation where $f_z$ is a function of $z$ and $v_z$ through $E_z$ and $\psi$ is a function of $z$. The likelihood function, which drives the method, compares the stellar number density in the $z-v_z$ plane to the model prediction. The best-fit potential is the one in which contours of constant $E_z$ coincide with contours of constant number density. The method has several advantages over other approaches. First, in contrast with methods based on the Jeans Equations, it works directly with the $z-v_z$ DF rather than its moments. Second, the DF and potential are inferred simultaneously rather than sequentially as in the approaches of \citealt{Kuijken1989a,Kuijken1989b,holmberg2000, holmberg2004}. Finally, evidence for disequilibrium such as the phase spiral emerge from the residuals of the model.

We also use this work to introduce the RLDF as a parametric model for $f_z$. In a sense, the RLDF serves as an alternative to models with thin and thick disc components. In fact, it has one fewer parameter than a model with two isothermal components. Since the RLDF can be written as the integral sum of isothermal components, it provides a simple mathematical model for the continuous mono-abundance sub-population proposal of \citet{bovy2012a, bovy2012b, bovy2013}.

From our analysis of a sample of GDR2 giants, we inferred the vertical potential, force, and density for $|z| \lesssim 1.5\,{\rm kpc}$. Our results were in general agreement with those found in the literature though our inferred value for the total mid-plane density was below the published values. A reanalysis using stars with relatively low radial energy (i.e., the "cold" population) yielded a value closer to the ones from the literature. 
We also calculated the vertical temperature distribution (differential surface density as a function of vertical velocity dispersion) and found that the form was similar to the corresponding distribution for SDSS/SEGUE G dwarfs from \citet{bovy2012b}. Finally, we viewed the residuals of the DF in the frequency-angle plane and found that the phase spiral mapped to a straight line whose slope yielded an estimate of $\sim540\,{\rm Myr}$ for the time since the event that perturbed the disc.

We conclude by mentioning two ways in which the model can be improved. The first is to tackle its most significant shortcoming, namely the use of the 1D approximation. To do so, one can model the DF by a three-integral DF from \citet{kuijken1995} or the quasi-isothermal model from \citet{binney2010, binney2011}. We note that in either case, we can replace the isothermal factor with an RLDF one. 
The analysis of RAVE data in \citet{piffl2014}, for example, adopts a quasi-isothermal DF and a global model for the potential, though their likelihood function involves velocity histograms (i.e., moments of the DF) due to the expensive computational cost of a likelihood function based on the full DF.
The second improvement is to incorporate disequilibrium features such as the phase spiral into the model. So long these features are {\it kinematic}, that is, do not perturb the gravitational field themselves, their structure will reflect the underlying potential. The idea is to model the equilibrium and disequilibrium components simultaneously. In the case of phase spirals, one might then model the number counts directly in frequency-angle space. The best-fit potential is then the one that maps contours of the equilibrium component to vertical lines and the phase spiral to a straight diagonal line.

\section*{Note added}

After submitting this work, we learned of the paper by \citet{Widmark2021} 
who inferred the local vertical potential by fitting the $z-v_z$ phase spirals while ignoring the bulk of the $z-v_z$ DF. 
Their method complements ours since we fit the bulk as an equilibrium distribution with the spirals emerging as residuals of the model. In our method, the spirals provide a consistency check in that they map to straight lines in the $\Omega(E_z)-\theta$ plane.

\section*{Acknowledgements}

It is a pleasure to thank Morgan Bennett, Jo Bovy, Sukanya Chakrabarti, Keir Darling, Zhao-Yu Li, Dan Foreman-Mackey, Kathryn Johnston, and Chervin Laporte for useful conversations. We also thank the referee for their insightful comments and useful suggestions.
We acknowledge the financial support of the Natural Sciences and Engineering Research Council of Canada.  
We also acknowledge funding from the Canada First Research Excellence Fund through the Arthur B. McDonald Canadian Astroparticle Physics Institute.
Lawrence M.Widrow is grateful to the Kavli Institute for Theoretical Physics at the University of California, Santa Barbara for providing a stimulating environment during a 2019 program on galactic dynamics. His research at the KITP was supported by the National Science Foundation under Grant No. NSF PHY-1748958.

\section*{Data availability}

The \textit{Gaia} Data Release 2 is available at the following website: https://gea.esac.esa.int/archive/. All other data used for our work is available through the links posted in the footnotes where necessary.



\bibliographystyle{mnras}
\bibliography{VerticalPotential_202102} 


\bsp	
\label{lastpage}
\end{document}